\documentclass[iop]{emulateapj}
\usepackage{epsfig}
\usepackage{natbib}
\usepackage{amssymb}
\usepackage{amsbsy}
\usepackage{natbib}
\usepackage{subfigure}
\usepackage[mathcal]{euscript}
\usepackage{float}
\usepackage{changepage}
\usepackage{graphicx}
\usepackage{tabularx}
\bibpunct[ ]{(}{)}{;}{a}{}{,}

\newcommand{\kepler}{{\it Kepler}}
\newcommand{\kms}{km~s$^{-1}$}
\begin{document}
\title{Testing Angular Momentum Transport and Wind Loss in
Intermediate Mass Core Helium Burning Stars}
\author{
Jamie Tayar\altaffilmark{1,2}, 
Marc H.~Pinsonneault\altaffilmark{1}
}
\altaffiltext{1}{Department of Astronomy, Ohio State University, 140 W 18th Ave, OH 43210, USA}
\altaffiltext{2}{Email: tayar@astronomy.ohio-state.edu}

\begin{abstract}

Stars between two and three solar masses rotate rapidly on the main sequence, and their rotation rates in the core helium burning (secondary clump) phase can therefore be used to test models of angular momentum loss used for gyrochronology in a new regime. Because both their core and surface rotation rates can be measured, these stars can also be used to set strong constraints on angular momentum transport inside stars. We find that they are rotating slower than angular momentum conservation and rigid rotation would predict. Our results are insensitive to the degree of core-envelope coupling because of the small moment of inertia of the radiative core. We discuss two possible mechanisms for slowing down the surfaces of these stars: (1) substantial angular momentum loss, and (2) radial differential rotation in the surface convection zone. Modern angular momentum loss prescriptions used for solar-type stars predict secondary clump surface rotation rates in much better agreement with the data than prior variants used in the literature, and we argue that such enhanced loss is required to understand the combination of core and surface rotation rates. However, we find that the assumed radial differential rotation profile in convective regions has a strong impact on the predicted surface rotation rates, and that a combination of enhanced loss and radial differential rotation in the surface convection zone is also consistent with the data. We discuss future tests that can quantify the impact of both phenomena. Current data tentatively suggests that some combination of the two processes fits the data better than either one alone.

\end{abstract}
 
\section{Introduction}
\setcounter{footnote}{0}
Real stars rotate, and rotation can have profound consequences for stellar structure and evolution.  Despite this, rotation is frequently ignored in stellar models, or treated in a highly simplified fashion.  The main culprit is the complex physics governing angular momentum evolution.  Stellar evolution naturally generates strong internal shears, especially in evolved stars with rapidly contracting cores and expanding envelopes.  Angular momentum can then be carried by convection-driven waves, Reynolds stresses from internal magnetic fields, and via large-scale circulation currents and weak turbulence driven by shears or instabilities.  It is not a priori obvious which of these mechanisms is dominant.  As a result, a wide range of internal rotation profiles could in principle exist; in turn, this permits a wide range of mixing rates and structural effects.  Adding rotation therefore requires the consideration of a number of phenomena traditionally not included in stellar models.  Empirical guidance is thus essential for progress, but historically the constraints on internal rotation have been sparse; in evolved stars, even surface rotation rates have been difficult to infer.
 
With the advent of large time domain and spectroscopic surveys, however, the observational landscape has been radically transformed.  There now exist hundreds of measurements of core rotation rates of evolved stars. Core rotation rates can be measured because the rotationally-split gravity modes propagating in the core can couple with surface pressure modes at similar frequency to form mixed modes which are visible on the surface but contain information on the core rotation \citep{Beck2012}. Measurements by \citet{Mosser2012b} suggested that core rotation periods for first ascent giants are of order tens of days and core rotation rates for helium burning stars are of order hundreds of days. 

As stars expand into red giants, their surface rotation must slow down to conserve angular momentum. Historically, this has made measuring surface rotation rates difficult. However, as the number of evolved stars monitored photometrically and measured spectroscopically has increased, it has become clear that in some of the more rapidly rotating giants, surface rotation measurements are possible. Specifically, surface rotation rates come from measurements of photometric modulation due to star spots \citep{Ceillier2017}, velocity broadening of spectral lines that can be measured in high resolution spectra \citep{Massarotti2008, Tayar2015}, and Doppler-like splittings of the stellar surface pulsations \citep{Deheuvels2015}.

In this paper, we constrain the degree of differential rotation and angular momentum loss in evolved stars. For this purpose, we chose to focus on intermediate mass, core helium burning stars, a sample which overlaps with the secondary clump identified by \citet{Girardi1998}. Isochrone fitting of binaries in this mass range indicates minimal mass loss in such stars \citep{Torres2015} consistent with the short timescale in which they cross the Hertzsprung gap. 
Main sequence rotation distributions have been measured for such stars \citep{ZorecRoyer2012} and because most of these intermediate mass stars do not undergo a helium flash, these rotational distributions can be smoothly forward modeled onto the secondary clump. The wide range of rotation rates on the main sequence is expected to produce relatively rapid rotation even in the core helium burning phase, producing seismically detectable core and envelope rotation as well as measurable spot modulation periods and velocity broadenings. 
Additionally, we find stars in this mass range particularly interesting because, while they are low enough mass to form a substantial fraction of the \kepler\ sample, their main sequence evolution is more similar to that of higher mass stars. We therefore hope to use intermediate mass stars as a bridge to understand the important processes affecting the rotation of massive stars, which can have substantial impacts on, for example, nucleosythetic processes in such stars and the energy budget of supernova progenitors.

Except for mass-dependent mass loss, the main sequence rotational evolution of all stars above about 1.3 M$_\sun$ \citep[the Kraft break,][]{Kraft1967} is thought to be similar.  These stars are born with a wide, somewhat mass-dependent range of rotation rates \citep{Gray1982, Finkenzeller1985,Alecian2013}. In stars without strong primordial magnetic fields or tidally interacting companions \citep[Ap and Am stars respectively,][]{Hubrig2000,Debernardi2000}, the lack of a deep surface convection zone means that such stars do not lose substantial angular momentum to a magnetized wind on the main sequence and their range of rotation rates persists to the end of the main sequence \citep{DurneyLatour1978}.

The rotational evolution of such stars after they develop a surface convection zone on the post-main sequence is much less constrained. At that point, one must begin to consider not only the direct effects of structural evolution, but also the possibility of non-rigid rotation profiles. Structurally, while the slow evolution and long lifetime on the main sequence make the assumption that the whole star rotates as a solid body seem reasonable, rigid rotation during rapid post-main-sequence evolution is substantially less likely. 
One must therefore consider the possibility of decoupling between the shrinking core and the growing envelope, as well as the possibility of radial differential rotation in both the radiative zone \citep[e.g.][]{Deheuvels2015} and the convective envelope \citep[e.g.][]{KissinThompson2015}.  Because red giants have very deep surface convection zones, differential rotation in such regions can have a very strong impact on their expected surface rotation rates.  As an illustration, \citet{Peterson1983} detected rapid rotation in blue horizontal branch stars, and their main sequence precursors are very slow rotators; furthermore, there is strong mass loss on the upper red giant branch.  \citet{Pinsonneault1991} and \citet{Sills2000} concluded that this combination required strong differential rotation with depth in the surface convection zones of luminous red giants, and probably differential rotation with depth in their radiative cores as well.  The problem is further complicated by the feedback of the rotation profile on the stellar structure \citep{MaederMeynet2000}.

In addition to varied rotation profiles, to understand rotational evolution one must also consider the effects of loss, which can be strongly mass dependent.  On the giant branch, mass loss is usually parameterized by a scaling which includes dependencies on the star's luminosity, gravity, and radius \citep{Reimers1975}. 
In low-mass main sequence stars, the effects of a magnetized wind are usually considered \citep{Kawaler1988} but an explicit parameterization of mass loss is rarely used. The \citet{Kawaler1988} formulation predicts torques that are a weak function of stellar radius, a conclusion challenged by \citet{ReinersMohanty2012} on the basis of how magnetic fields were scaled relative to the solar case.  Solutions for magnetized stellar winds \citep{Matt2012}, rather than general scalings, were then found to predict a much stronger dependence of the torque on stellar properties, especially radius.  Wind laws using this general approach are now being used for gyrochronology and angular momentum evolution models of low mass stars \citep{vanSadersPinsonneault2013, GalletBouvier2013, LanzafameSpada2015, Matt2015}. It is, however, difficult to distinguish between such models directly using only low mass stars \citep[see][for a recent example]{Somers2017}.  Since these same angular momentum loss laws form the basis for gyrochronology, changing their form could alter the inferred ages for many stars, especially those most different from the sun.

In this paper, we collect the available data on the surface rotation rates in the secondary clump from a variety of methods (Section \ref{sec:methods}). We also construct a theoretical framework for interpreting that data in the context of structural evolution, angular momentum loss, and radial differential rotation (Section \ref{sec:methods}), discuss the predicted rotation trends with mass and radius (Section \ref{sec:trends}) and compare the predicted and observed rotation distributions for each of our model cases (Section \ref{sec:Distributions}). While our work sets some bounds on core rotation, which we discuss, we will demonstrate that the core coupling has only a minor impact on the predicted surface rates. We therefore postpone detailed discussion on the measured and predicted rates of core rotation as a function of mass and surface gravity to a companion paper that also includes a carefully selected set of new core rotation measurements (Tayar et al., in prep).

\section{Methods}\label{sec:methods}

Our goal is to construct self-consistent evolutionary models including rotation for intermediate mass stars, with the goal of evaluating different scenarios in the core-helium burning (secondary clump) phase.  There are a number of ingredients that must be addressed in this exercise, starting with the classical ingredients of our models.  We are interested in the difference between main sequence and evolved stars, so we need to ensure that our theoretical models are consistent with the observed locus of evolved stars in the HR diagram; ideally, our models would also be consistent with observed main sequence constraints.  We also need a proper set of initial conditions (in particular, initial rotation rates) and we consider different scenarios for mass and angular momentum loss.  Our work in this paper is primarily focused on interpreting surface rotation measurements, but different scenarios for internal angular momentum transport in radiative regions can impact our results, and they need to be delineated.  Finally, we need to discuss a treatment of the rotation profile imposed in convective regimes.  On the data side, we need to define the mass regime of interest, discuss how our data samples were constructed, and consider the (significant) observational biases that can impact surface measurements.  We begin in Section \ref{Data} with a discussion of our data, and follow in Sections \ref{sec:Models} and \ref{sec:RotPhysics} with the standard and rotational physics of the models, respectively.

\subsection{Post-Main-sequence Samples} \label{Data}
\begin{figure}[h]

 \centering
\subfigure{\includegraphics[width=0.44\textwidth, clip=true, trim= 0in 0.2in 2.3in 2.7in ]{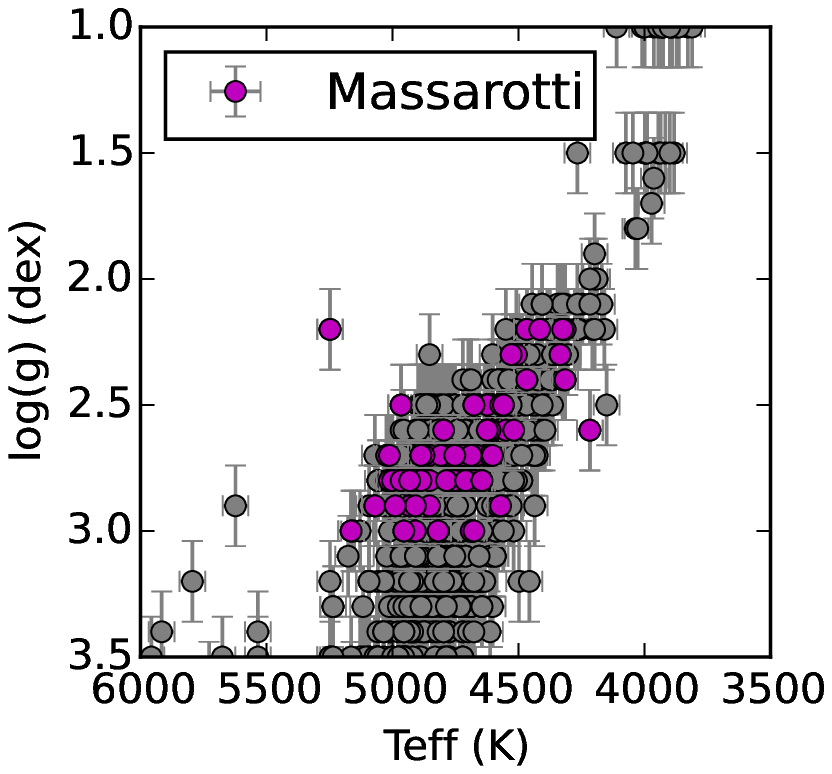}}
\subfigure{\includegraphics[width=0.44\textwidth, clip=true, trim= 0in 0.2in 2.3in 2.7in ]{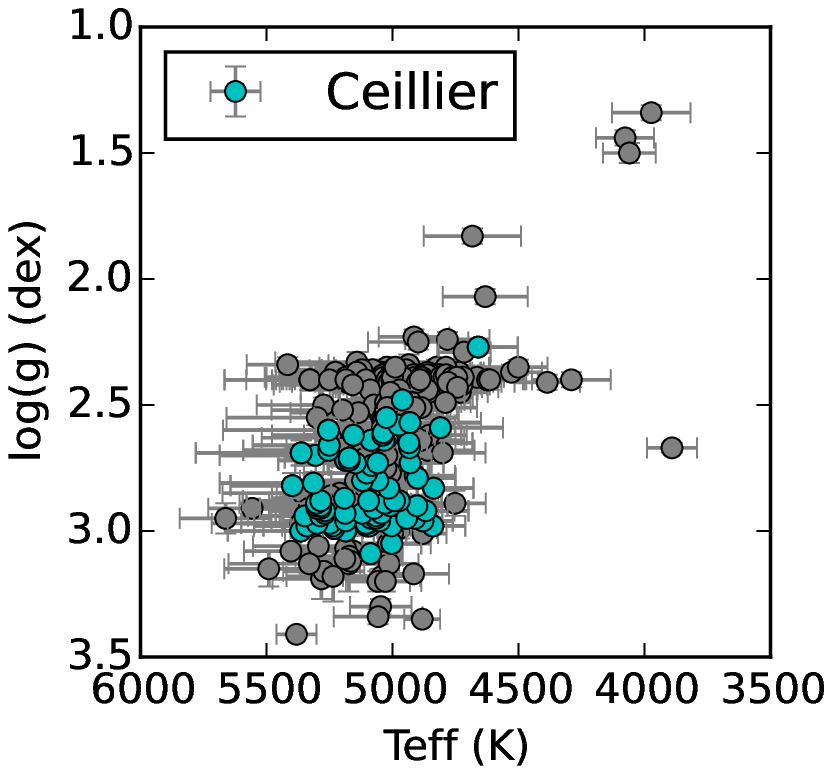}}

\caption{HR diagram position of the stars between two and three solar in the log(g) range of our sample compared to the stars in the \citet{Massarotti2008} (top) and \citet{Ceillier2017} (bottom). Stars that fall within our mass and gravity cuts are colored, other stars in the sample with measured rotation rates are shown in gray.}
\label{Fig:hrsample}
\end{figure}

In order to understand the physical mechanisms involved in determining stellar rotation rates, we wish to compare the distribution of rotation rates predicted by various models to actual measurements. For this comparison, we would ideally like a large, unbiased data set with minimal contamination by either binaries or lower mass stars, reliable measurements to arbitrarily low velocities and matched core and surface rotation rates. Because no such data set exists, we instead compare our model results with five different available data sets, each with its own strengths. The first is a volume limited sample from \textit{Hipparcos} which has minimal selection effects but lacks precise mass information. We add to this several samples from the \textit{Kepler} field obtained using different techniques for measuring surface rotation including v sin(i), period, and oscillations. These targets have precise asteroseismic masses, but each is only sensitive to rotation in a limited domain. This can cause significant sample selection effects. As a result, for our analysis, we will compare to multiple data sets. For each sample, we use only stars between 2.0 and 3.0 M$_\sun$ with log(g) between 3.1 and 2.2 dex (see Figure \ref{Fig:hrsample} for the temperatures and gravities in two of our samples).

\begin{figure*}[t]
\begin{minipage}{\textwidth}

 \centering
\subfigure{\includegraphics[width=0.45\textwidth] {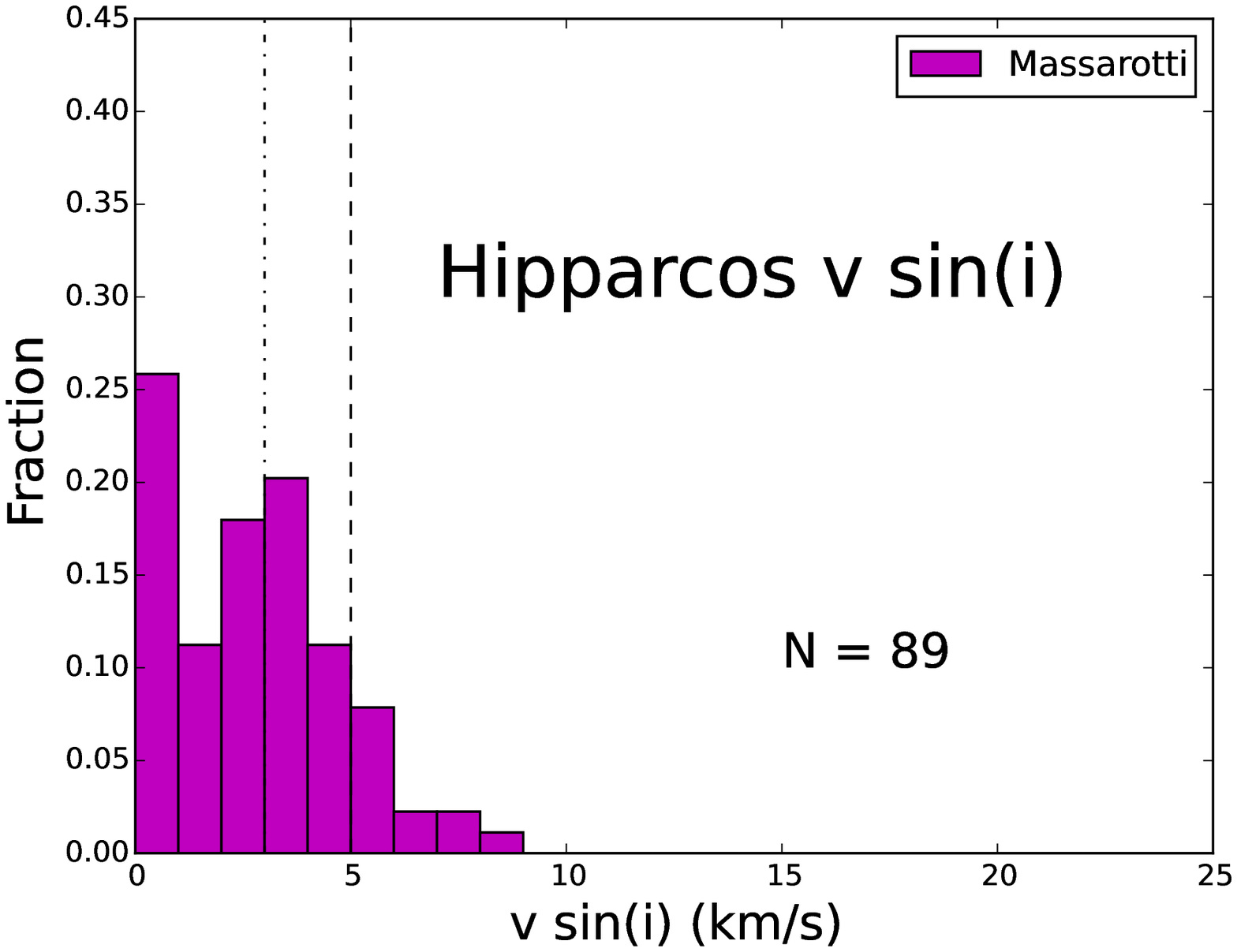}}
\subfigure{\includegraphics[width=0.45\textwidth ] {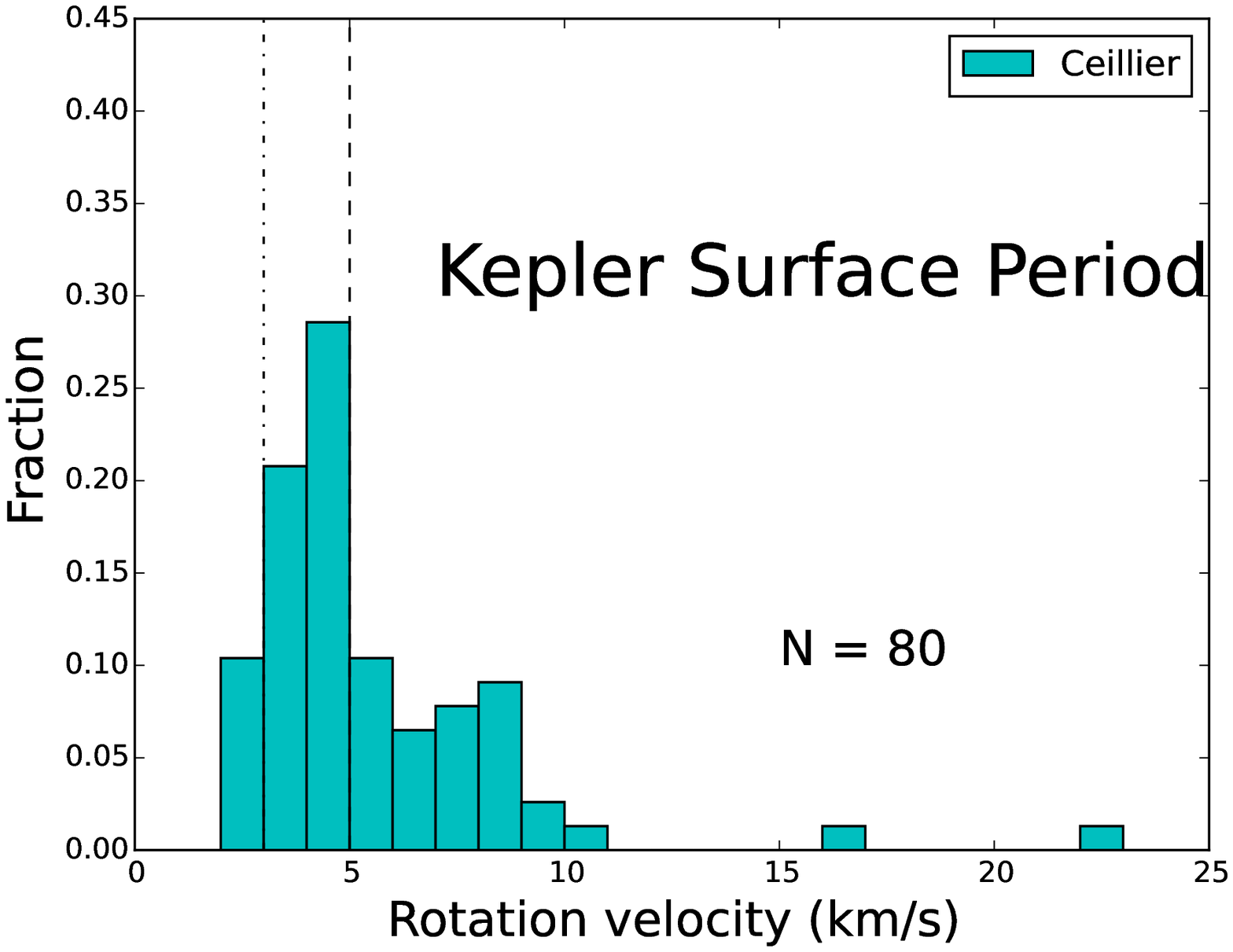}}
\subfigure{\includegraphics[width=0.45\textwidth ] {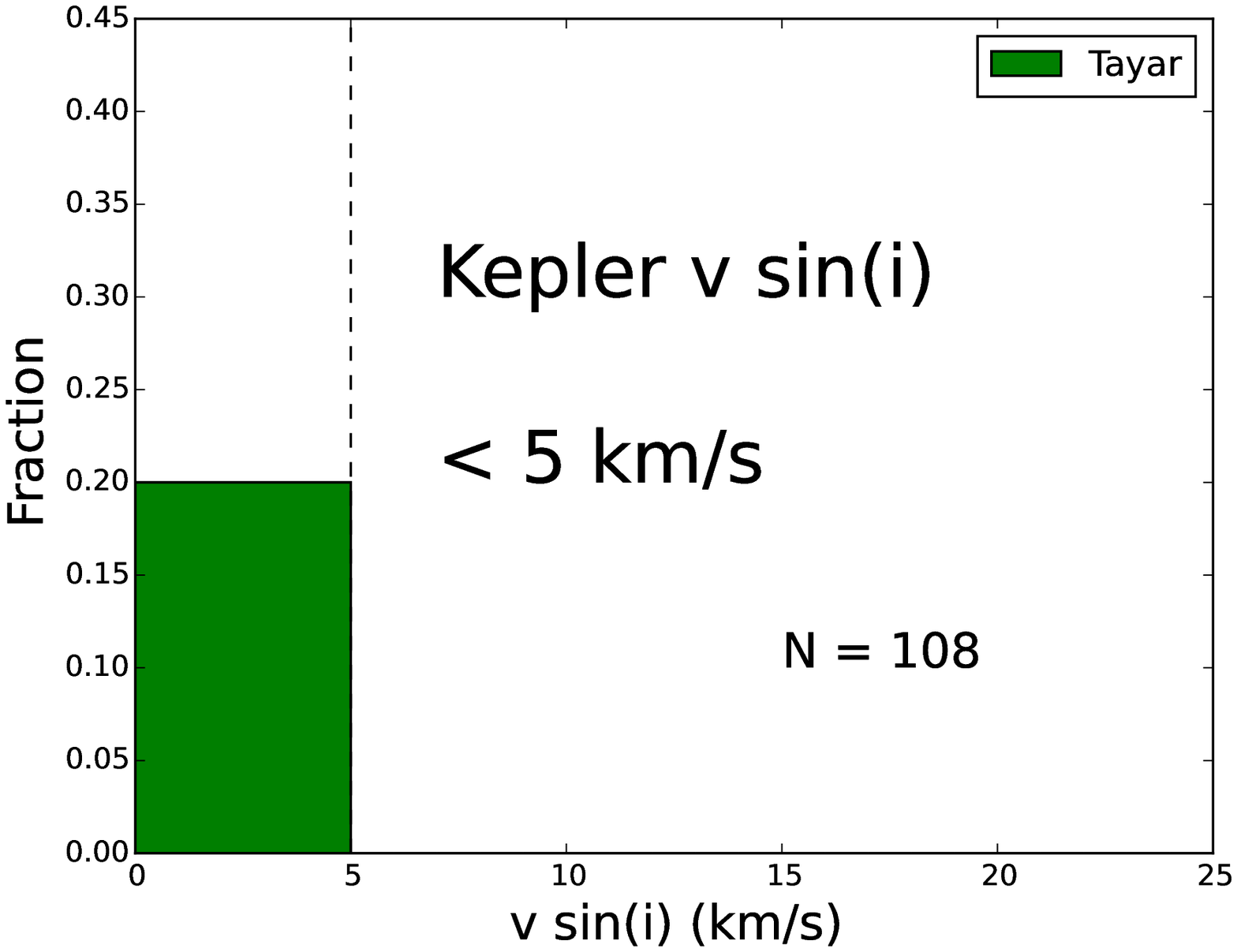}}
\subfigure{\includegraphics[width=0.45\textwidth ] {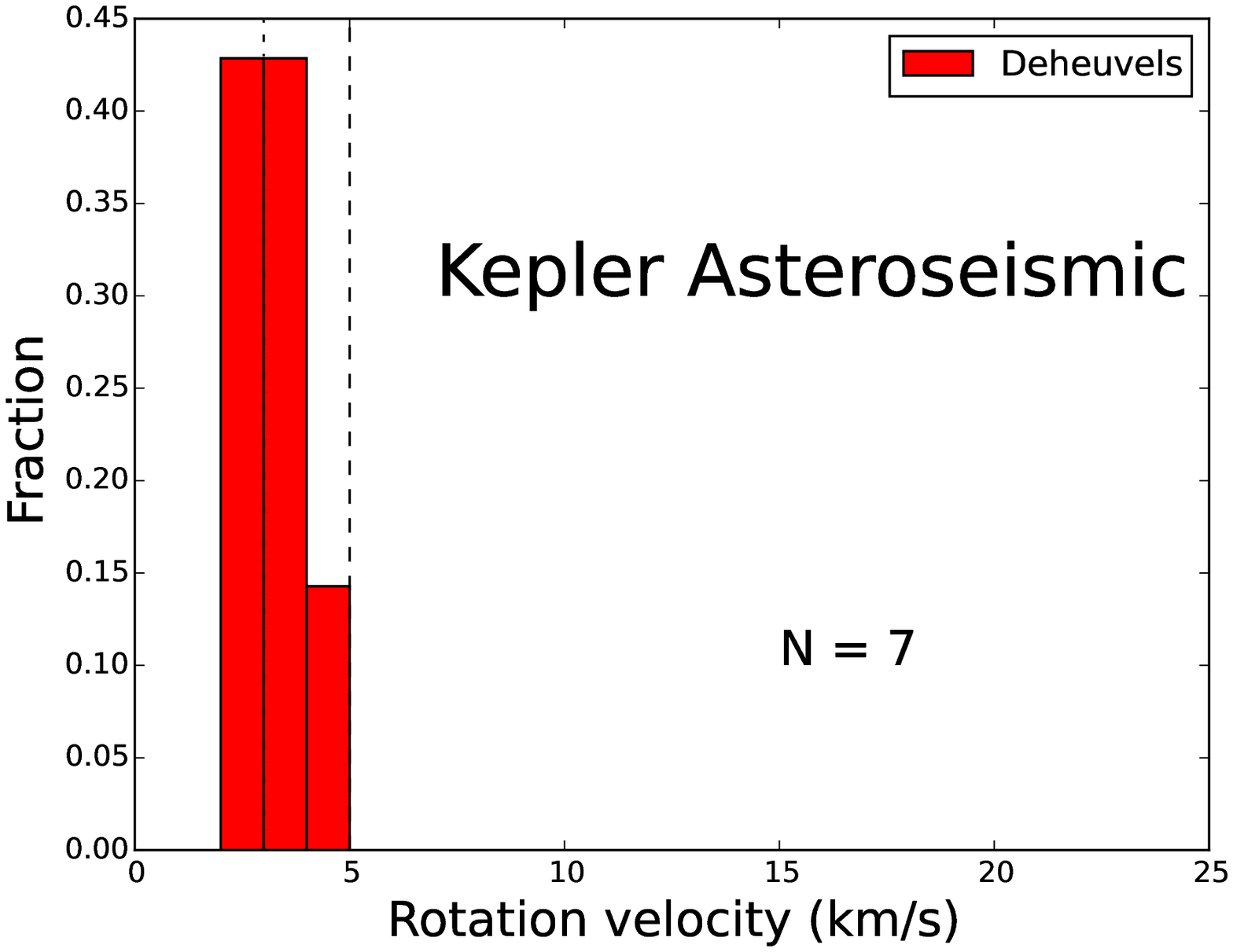}}
\caption{Surface rotation rate distributions for our four samples \citep{Massarotti2008, Tayar2015, Deheuvels2015, Ceillier2017} of surface rotation rates. Note that spectroscopic rotation rates include the inclination angle ambiguity while period and frequency measurements do not. The number of stars in each sample is listed in the lower right of each plot. Dashed vertical lines mark a cutoff at 5 \kms; dot-dashed vertical lines mark 3 \kms. Sample colors will be consistent through this work. 
}
\label{Fig:SurfaceFig}
\end{minipage}

\end{figure*}%

\begin{figure*}[t]

\begin{minipage}{\textwidth}
 \centering
\subfigure{\includegraphics[width=0.45\textwidth ]{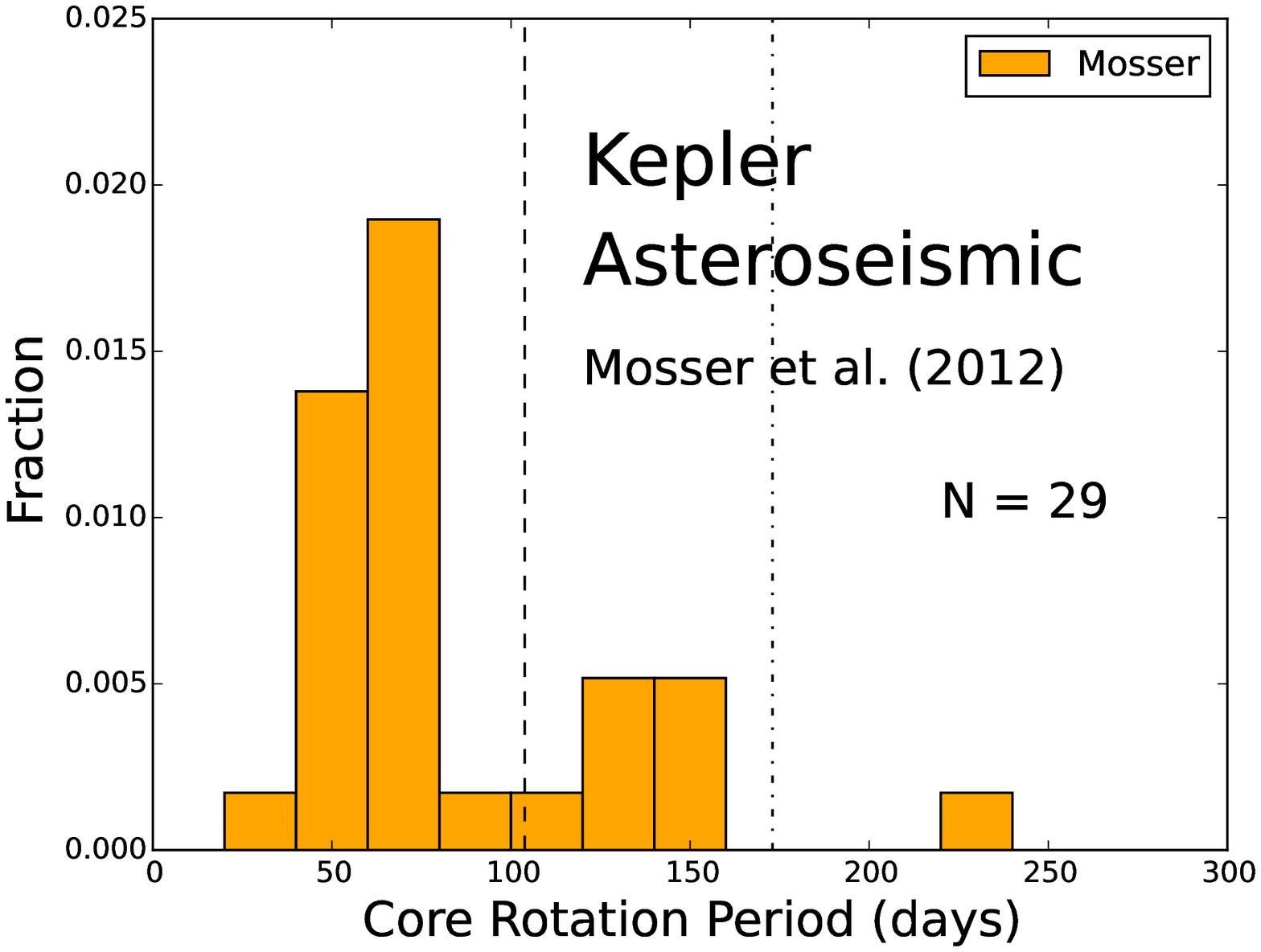}}
\subfigure{\includegraphics[width=0.45\textwidth ] {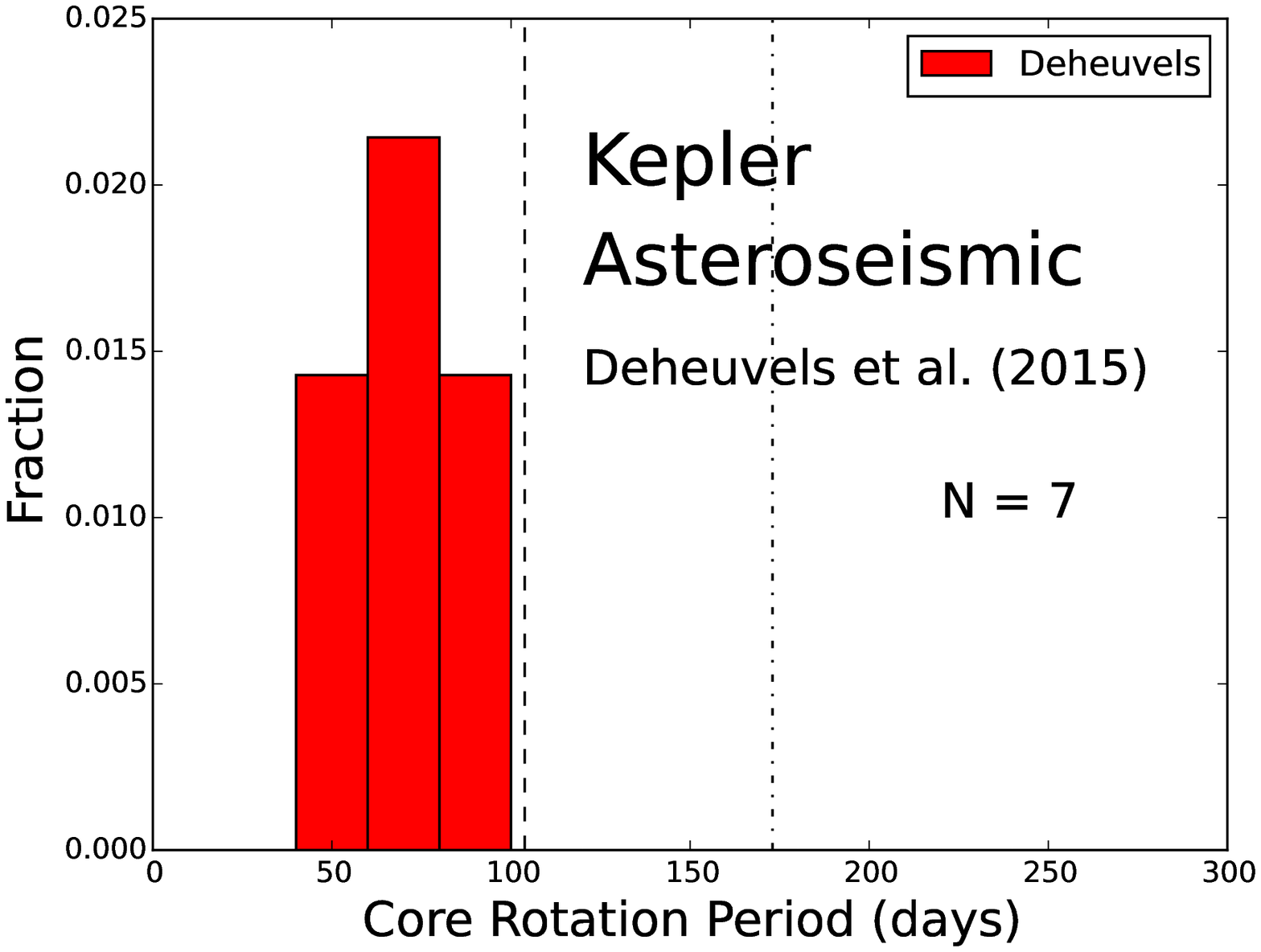}}
\caption{Core rotation period distributions for our two samples \citep{Deheuvels2015, Mosser2012b}. The number of stars in each sample is listed in the lower right of each plot. Dashed vertical lines mark a cutoff at 103.8 days (approximately 5 \kms\ at the average radius of the Deheuvels et al. sample); dot-dashed vertical lines mark 173.1 days (approximately 3 \kms\ at the average radius of the Deheuvels et al. sample). Sample colors will be consistent through this work.}
\label{Fig:CoresFig}
\end{minipage}

\end{figure*}

\twocolumngrid

\subsubsection{Hipparcos Field Stars} 
One of the largest and most unbiased samples available is the analysis of the \textit{Hipparcos} sample of giants within 100 parsecs by \citet{Massarotti2008}. In addition to being essentially volume limited, extensive work has been done to remove binary stars. Line broadenings for each star were computed by cross-correlating with a grid of template spectra at various rotational broadenings and fixed microturbulence and macroturbulence values. These line broadenings were then converted to rotational velocities assuming temperature and luminosity dependent macroturbulence. Rotation rates are quoted down to less than 1 \kms; these stars do not have measured core rotation rates. However, this sample relies on color based estimates of effective temperature using available literature data, and assuming a metallicity if one was not available. The surface gravities are computed in a heterogeneous fashion, either taken from the literature or based on comparison with evolutionary tracks.  

Because mass determinations for this sample involve combining the \textit{Hipparcos} distances and fluxes with the spectroscopic temperatures and gravities, the mass uncertainties are large. Comparison with a more recent analysis by \citet{Feuillet2016} indicates that the standard deviations for the measurements are 51 K, 0.16 dex and 0.83 M$_\sun$, which we take as the error on the values.  To ensure consistency with our other data, we also have to put the mass inferences on a common system. When we compare stars calculated to be between two and three solar masses by \citet{Massarotti2008} to the more recent results of \citet{Feuillet2016} for these same stars, we find that the Massarotti measurements are on average 150 K cooler, and 0.13 M$_\sun$ larger, although the surface gravities agree on average. To account for the mass shift, we adjust the computed masses down by 0.13 M$_\sun$ before selecting our sample.  Because we do not use the temperature in our interpolation, and because of the uncertainty in the true temperature of red giants, we choose not to correct the temperatures for these stars. In our final sample, there are 89 stars; 87 percent have v sin (i) $<$ 5 \kms\ and 55 percent have velocities less than 3 \kms, with a maximum of 8.1 \kms.

\subsubsection{Kepler Surface Rotation}\label{Data:Tayar} 
With data of sufficient quality, asteroseismology can be used to measure surface rotation rates, using p-dominated rather than g-dominated modes. \citet{Deheuvels2015} published a sample of stars with measured masses, radii, and core rotation rates with seismic envelope rotation rates that are reliable to very low velocities ($<$ 3 \kms). Because the inclination is a seismic observable, they have no sin(i) ambiguity. However, this sample is quite small, containing only 7 stars, and it is not immediately clear that the measurements of envelope rotation deduced from the p mode splittings give the same information as a surface rotation measurement would, especially in the case of strong differential rotation \citep[but see e.g.][for some encouraging results on the main sequence and red giant branch, respectively]{Gizon2013, Beck2017}. Of the 7 stars in the sample, all 7 have inferred surface velocities less than 5 \kms\ and 3 (43\%) have inferred surface velocities less than 3 \kms. The core rotation periods for these stars are all shorter than 103.8 days, the rotation period which corresponds to 5 \kms\ at the average radius of the sample, indicating that the cores are rotating faster than their envelopes.

The APOGEE-\kepler\ red giant sample \citep{Pinsonneault2014} combines large numbers of spectra with asteroseismic analysis. We therefore add to these samples the \citet{Tayar2015} spectroscopic analysis of the APOGEE-\kepler\ sample of giants, which have seismic masses and radii. However, the resolution of the APOGEE spectrograph prevents measurements of velocity broadenings less than 5 \kms. There are no core rotation rates publicly available for this sample. By comparison with field data \citep[e.g.][]{Carlberg2011}, \citet{Tayar2015} found that the \textit{Kepler} asteroseismic sample was biased against rapidly rotating stars. However, given the size of the sample (108 stars), the lack of even a single detection with a v sin i above 5 \kms\ is quite constraining. We note that the open cluster sample from \citet{Carlberg2014} designed to study stars in a similar mass range to ours has a similar distribution as the \citet{Tayar2015} and \citet{Massarotti2008} samples.

Finally, we add to these velocity measurements a sample of rotation period measurements from \citet{Ceillier2017} for the \kepler\ active oscillating giant sample. We emphasize that while the requirement of oscillations likely biases the sample against rapid rotators, the period range searched and the requirement of significant activity likely selects for the fastest rotating stars. These stars have seismic masses and radii which allow the period measurements to be converted to velocities without an inclination degeneracy. This particular period measurement technique has been shown to recover periods in about 80 percent of stars with about 90 percent of recovered periods being reliable \citep{Aigrain2015}. 
This sample of 80 stars has surface velocities between 2.4 and 88 \kms, although we note that the few velocities on the very high end of this sample are likely to be spurious detections of the rotation periods of a nearby, smaller star \citep{Ceillier2017}. 58 percent of stars have velocities below 5 \kms\ and only 10 \% have velocities below 3 \kms. 

The rotation distributions for the surfaces of all four samples can be seen in Figure \ref{Fig:SurfaceFig}. Three of the distributions are consistent with a peak around 3 \kms, the \citet{Ceillier2017} sample seems to peak closer to 5 \kms. For our analysis, we will therefore compare both the \citet{Massarotti2008} and \citet{Ceillier2017} samples to our predictions, and as we will show, we make similar inferences from the two samples, implying that the selection effects are unimportant for our purpose.

\subsubsection{Kepler Core Rotation} 
The final sample we consider in this work is the \citet{Mosser2012b} sample of core rotation periods. This sample contains 29 stars above 2 $M_\sun$ with seismically measured masses and radii whose core rotation periods have been inferred from the Doppler-like splittings of the core influenced mixed modes. While this is a relatively small subset of the full Kepler sample and could therefore contain significant selection effects, it is nevertheless the best sample to date for inferring the distribution of core rotation rates for this mass range. We find that 28\% of stars have core rotation periods longer than 103.8 days (an approximate analog of 5 \kms\ using the average radius of the Deheuvels et al. 2015 sample) and only 3\% of stars have rotation periods longer than 173.1 days (approximately equivalent to 3 \kms). This would seem to indicate that the cores of the stars in this sample are rotating faster on average than the surface rotation rates measured spectroscopically, and are more consistent with the distribution of surface rotation periods measured photometrically. We show the rotation distributions for both of the core rotation samples in Figure \ref{Fig:CoresFig}.

\subsection{Theoretical Models-Standard Model Physics} \label{sec:Models}
\begin{figure}[h]

 \centering
\includegraphics[width=0.44\textwidth, clip=true, trim= 0in 0.2in 2.3in 2.7in ]{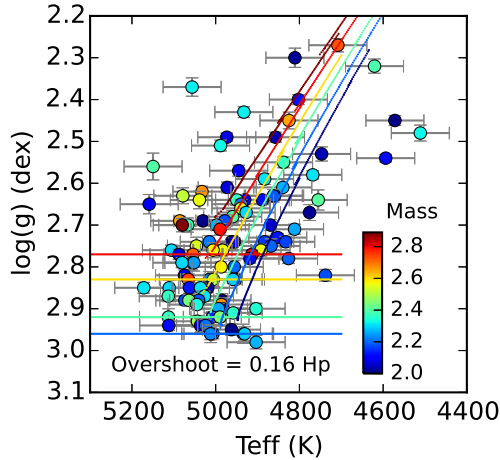}
\caption{HR diagram position of the stars in the \citet{Tayar2015} sample compared to the positions of stellar models with our chosen best fit overshoot. In each plot, the horizontal colored lines indicate the 90$^{th}$ percentile surface gravity of stars within 0.1 M$_\sun$ of each mass modeled. }
\label{Fig:Overshoot}
\end{figure}

The physics of our models are summarized in Table \ref{Table:physics}, and we show the comparison of our models and data on a Kiel diagram in Figure \ref{Fig:Overshoot}. Models were constructed using the Yale Rotating Evolution Code \citep[YREC,][contains some discussion of more recent upgrades to the input physics] {Pinsonneault1989, vanSadersPinsonneault2012}. We use a \citet{GrevesseSauval1998} mixture of heavy elements and solar metallicity (Z/X), \citet{Kurucz1997} atmospheres with pressure evaluated at $\tau = \frac{2}{3}$, OPAL high temperature opacity tables \citep{IglesiasRogers1996}, \citet{Ferguson2005} low temperature opacity tables, and the updated OPAL equation of state \citep{Rogers1996, RogersNayfonov2002}. We do not include the effects of diffusion or semi-convection \citep[see eg.][]{Kippy}. 
We use a mixing length and helium taken from \citet{Tayar2017} which have been calibrated to reproduce the temperature of the first ascent giant branch as a function of metallicity for the combined APOGEE-\textit{Kepler} sample (Pinsonneault et al, in prep.). 

We will use surface gravity and mass to map our rotation detections to models. Therefore, one variable of importance to this work that was not calibrated by \citet{Tayar2017} is the overshoot of the core, as the core overshoot determines the minimum core size and thus the maximum surface gravity of core helium burning stars. Work on eclipsing binaries \citep{ClaretTorres2016} and asteroseismology \citep{Aerts2013} has suggested that core overshooting parameters of between 0.1 and 0.2 pressure scale heights are most consistent with observations. For this analysis, we are most concerned with models that have the correct moment of inertia, and we therefore want to ensure that the maximum surface gravities populated by the data exist in the core helium burning phase of our models. 

We choose a benchmark 2.4 M$_\sun$ model to calibrate the overshoot parameter, which is just above the average mass (2.28 M$_\sun$) of this sample, and the first mass for which all models run smoothly through the core helium burning phase. We ran models with overshoot parameters of 0, 0.1, and 0.2 pressure scale heights and found the surface gravity of the Zero Age Horizontal Branch (ZAHB) for each model. We then take the stars between 2.3 and 2.5 M$_\sun$ in the \citet{Tayar2015} sample\footnote{As our mixing length is calibrated on Sloan Data Release 13 \citep{DR13} temperatures, we compare the models to the data using these DR13 temperatures. As we use only the log(g) values for detailed analysis later, this does not impact our analysis.}  (see Section \ref{Data:Tayar}), and compute the surface gravity of the 90th percentile star in that mass range. We then interpolate to find the overshoot parameter such that the surface gravity of the ZAHB would equal the surface gravity of the 90$^{th}$ percentile star. Choosing a different mass bin or using the 95$^{th}$ percentile would change the overshoot parameter by less than 0.03 H$_{\rm p}$. To prevent unphysically large overshooting in models with small cores, we also enforce that the core overshooting cannot be larger than 0.15 times the radius of the core.

\begin{table}[htbp]
\caption{Summary of the input physics used in our models.}
\begin{tabularx} {.48\textwidth}{>{\raggedright\arraybackslash}X|>{\raggedright\arraybackslash}X} 
\hline\hline
Parameter & YREC \\ \hline
Atmosphere & \citet{Kurucz1997}\\ 
$\alpha$-enhancement & No \\ 
Convetive Overshoot & 0.16H$_{\rm p}$ or 0.15 R$_{\rm core}$  \\ 
Diffusion & No  \\ 
Equation of State & OPAL+SCV \\ 
High Temperature Opacities& OPAL \\ 
Low Temperature Opacities & \citet{Ferguson2005}  \\ 
Mixing Length & 1.90 \citep{Tayar2017}  \\ 
Mixture and Solar Z/X & \citet{GrevesseSauval1998} \\ 
Nuclear Reaction Rates & \citet{Adelberger2011} \\ 
Weak Screening & \citet{Salpeter1954}  \\ 
Solar X & 0.709306  \\ 
Solar Y & 0.272683  \\ 
Solar Z & 0.0179471  \\ 
Mass Range & 2.0 M$_\sun$ to 3.0 M$_\sun$ \\\hline
\end{tabularx}
\label{Table:physics}
\end{table}

\subsection{Theoretical Models- Rotational Physics}\label{sec:RotPhysics}
The key parameters determining the rotation rate of a star include the initial rotation rate, mass loss, angular momentum loss, radial rotation profile, and presence or absence of mechanisms that redistribute angular momentum. We discuss below the rational for each of our rotation choices, and summarize our various cases in Table \ref{Table:rotation}.

\begin{table*}[t]
\caption{Summary of the rotation cases we consider}
\begin{tabular}{rlll}

\multicolumn{1}{l}{Cases} & Loss & Radial Rotation Profile &  \\ 
\multicolumn{1}{l}{} &  & Core & Surface Convection Zone \\ 
1 & Kawaler & Solid Body Central Convective and Radiative Zone & Solid Body at Core Rate \\ 
2 & Kawaler  & Solid Body Central Convective and Radiative Zone & Differentially Rotating Convective Zone \\ 
3 & PMM & Solid Body Central Convective and Radiative Zone & Solid Body at Core Rate \\ 
4 & PMM & Solid Body Central Convective and Radiative Zone & Differentially Rotating Convective Zone \\ 
\end{tabular}
\label{Table:rotation}
\end{table*}

\subsubsection{Initial Conditions} \label{ssec:ZR2012}
Rather than attempting to match the pre-main-sequence rotational evolution of our stars, we choose instead to use the measured distribution of rotation rates in main sequence turnoff field stars of the appropriate mass to determine our initial rotation conditions. We assume that solid body rotation is effectively enforced by the end of the main sequence \citep[reasonable according to the results in][]{Aerts2017}.
A distribution of main sequence rotation rates in constructed using the analytical rotation distributions of \citet{ZorecRoyer2012}, who measured projected velocity broadenings of \textit{Hipparcos} stars. Masses and ages of the observed stars were estimated by matching the observed temperatures and luminosities to evolutionary tracks from \citet{Schaller1992}. The rotation distributions were broadened and corrected for inclination effects and each mass range was fit as the sum of two Maxwellians, a slow component and a fast component. 

To construct the distributions for our mass range, we use the distribution computed for stars between 2.3 and 2.7 M$_\sun$, the middle of our range. Specifically, we draw our initial distribution of stars from two Maxwellians\footnote{$f(x)=(x-l)^2\sqrt{2/\pi}/\alpha^3 e^{(-(x-l)^2/(2\alpha^2))}$}, a slow distribution comprising 15 percent of the stars with a lag $l=0$ and a scale parameter $\alpha=28.28$ and a fast distribution comprising 85 percent of the stars with a lag $l=0$ and a scale parameter $\alpha=142.13$. We note that using a distribution determined for a slightly different mass range within our sample does not significantly effect our results. To avoid drawing extremely fast rotators from the tail of the distribution, where the rapid rotation rates are numerically challenging to simulate, we set all rotation rates larger than the maximum of our grid to 10 \kms\ less than that maximum. Given our loss proscriptions, where the loss depends on rotation rate, we would expect these extremely rapid rotators to converge down to the rotation rates we simulate relatively quickly and have a minimal effect on our predictions.

\citet{ZorecRoyer2012} inferred a slightly different rotation distribution when they excluded known chemically peculiar stars and binary members, two groups thought to rotate more slowly than the average star in this mass range \citep{Abt2009}. Such stars represent approximately 36\% of the stars in this mass range in the Zorec \& Royer sample. While we expect all close binaries in our sample to have merged, we expect that the descendants of single chemically peculiar stars will be present in our sample as an excess of slowly rotating stars. We therefore use the Zorec \& Royer distributions derived for the entire sample including the chemically peculiar stars even though it might slightly overestimate the number of slow rotators we expect. We also note that pulsational broadening of the lines can be misinterpreted as rotational broadening \citep{Aerts2014}, adding an additional uncertainty of up to tens of kilometers per second to each measurement. We return to the impact of these choices in Section \ref{DistCompare}.

\subsubsection{Thermal Wind Loss} \label{ssec:thermal}
Because the mass and angular momentum loss from a non-magnetized wind in the regime we consider is limited, we consider it separately from angular momentum loss through a magnetized wind and do not consider the feedback of mass loss on the stellar structure. We assume a \citet{Reimers1975} mass loss rate of $\dot{M}=1.27\times10^{-5}\eta M^{-1}L^{1.5}T_{eff}^{-2}$ M$_\sun$ yr$^{-1}$ with $\eta=0.4$ \citep{Bertelli1994, RenziniFusiPecci1988}, a value which likely represents an overestimate of loss in our stars \citep{Miglio2012}. We assume that mass is lost in a spherically symmetric thermal wind from the stellar surface, which causes angular momentum loss $\dot{J}=-\frac{2}{3}\omega R^2 \frac{dM}{dt}$ \citep{Mestel1968}.

\subsubsection{Magnetized Wind Loss} \label{ssec:Magnetized}
We first model post-main-sequence angular momentum loss through a magnetized wind using the standard solar calibrated \citet{Kawaler1988} proscription $\frac{dJ}{dt}=-K\omega^3(\frac{R}{R_\sun})^{0.5}(\frac{M}{M_\sun})^{-0.5}$  with $K=2.95 \times 10^{47}$ s up to a Rossby scaled critical rotation rate of  $\omega_{crit}=10\frac{\omega_\sun\tau_{\sun}}{\tau_{cz}}$. 
  Faster than that, angular momentum loss goes as $\frac{dJ}{dt}=-K\omega\omega_{crit}^2(\frac{R}{R_\sun})^{0.5}(\frac{M}{M_\sun})^{-0.5}$ \citep{Sills2000, Krishnamurthi1997}. 

The above wind law is a simplification of the general form of angular momentum loss derived by \citet{Mestel1968}. \citet{Kawaler1988} used results from \citet{LinskySaar1987} which indicated that R$^2$B$_0 \propto \Omega$ and substantially reduced the radius dependence of angular momentum loss. With the availability of newer data, it is now possible to fit for the full dependence of angular momentum loss. We therefore also consider a newer parameterization of angular momentum loss formulated by Pinsonneault, Matt and MacGregor, (hereafter referred to as PMM) and discussed in \citet{vanSadersPinsonneault2013} and \citet{Matt2012}. In this parameterization, $\frac{dJ}{dt}$ goes as $f_KK_M\omega(\frac{\omega\tau_{cz}}{\omega_\sun\tau_{cz,\sun}})^2$ with $f_K=3.78 \times 10^{47}$ and $\frac{K_M}{K_{M,\sun}}=c(\omega)(\frac{R}{R_\sun})^{3.1}(\frac{M}{M_\sun})^{-0.22}(\frac{L}{L_\sun})^{0.56}(\frac{P_{phot}}{P_{phot,\sun}})^{0.44}$ up to a Rossby scaled critical rotation rate $\omega_{crit}=12\frac{\omega_\sun\tau_{\sun}}{\tau_{cz}}$. Above that,$\frac{dJ}{dt}=f_KK_M\omega(\frac{\omega_{crit}}{\omega_\sun})^2$ .  Because of the larger radius dependence and the explicit dependence on the convective overturn timescale, we expect this loss law to have a much more noticeable effect on our evolved intermediate mass stars.

\subsubsection{Radial Rotation Profile}

In all of our models, we enforce rigid rotation in all regions below the surface convection zone. This also serves as a lower limit on the angular momentum content of the core assuming that the rotational profile must be monotonically declining in radius in radiative regions. Allowing differential rotation in radiative zones increases the angular momentum content locked up in the core and reduces the angular momentum content and thus the rotation rate of the envelope. 

We do, however, consider two rotation profiles for the surface convection zone.
The first enforces rigid rotation in all zones of the star at all times, analogous to the 'maximal coupling' case of \citet{TayarPinsonneault2013}. 
This profile has no dependence on radius, and therefore goes as R$^0$. We also consider the effects of allowing radial differential rotation in the surface convection zone, where the rotation rate goes as some non-zero power of the radius. The maximal amount of differential rotation we allow is R$^{-2}$, which represents constant specific angular momentum in the surface convection zone. In this case, the core is rotating more quickly than the rigid case would predict and the surface is rotating more slowly. We note that even if the core rotation rate is not fixed to the rotation rate of the base of the convection zone, angular momentum conservation {as the convective envelope shrinks in mass during the core helium burning phase}
keeps the two rotation rates very close. We therefore choose to fix the rotation rate of the core to the rotation rate of the base of the convection zone.

\section{Predicted Trends} \label{sec:trends}
In this section, we discuss the predictions of our various classes of models and compare them to the available measurements. We start in Section \ref{ssec3:standard} with a discussion of the structural evolution of these stars. In Section \ref{ssec3:thermal} we add the effects of a thermal wind, and in Section \ref{ssec3:magnetized} we consider a magnetized wind. We discuss the effect of allowing radial differential rotation in the convection zone in Section \ref{ssec3:diffrot}, and finish with a comparison between the predictions of each model and the available data in Section \ref{ssec3:observed}.

\begin{figure*}[t]
\begin{minipage}{\textwidth}
 \centering
\subfigure{\includegraphics[width=0.7\textwidth, clip=true, trim= 0.1in 0.3in 0.3in 0.5in ]{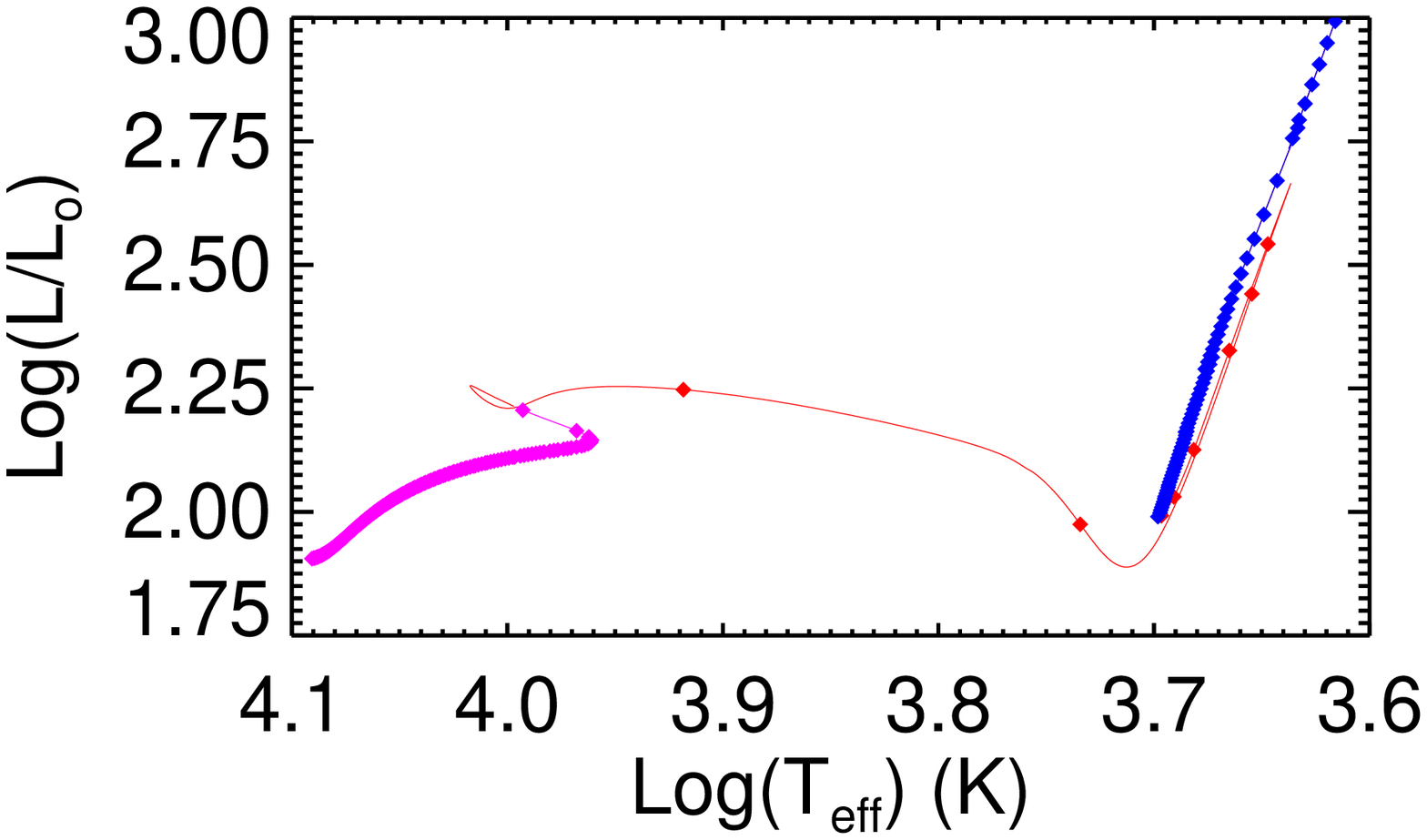}}\subfigure{\llap{\raisebox{4cm}{\includegraphics[height=2.4cm,clip=true, trim=-10.cm 0.8cm -8cm 1.4cm]{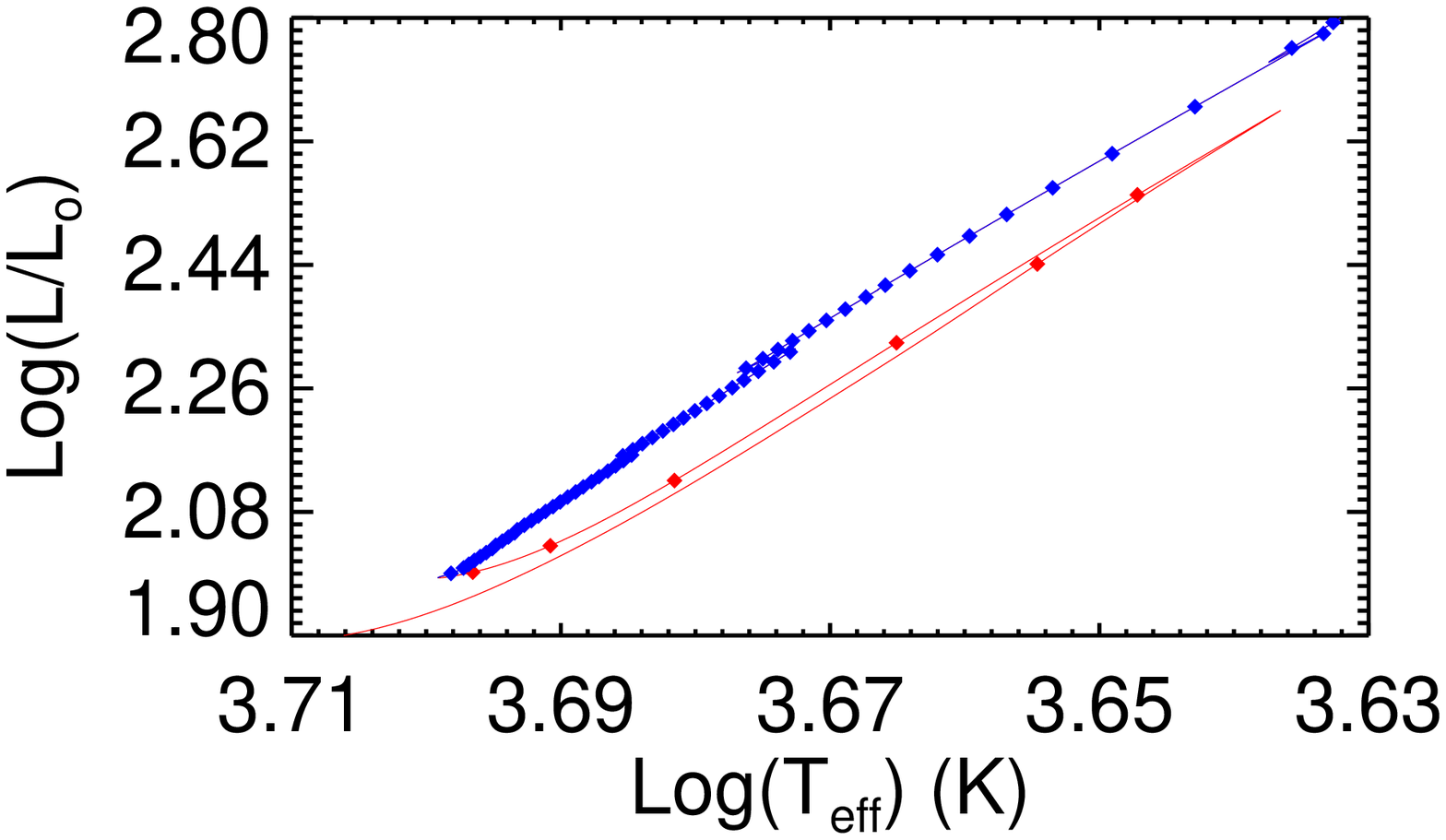}}}}

\subfigure{\includegraphics[width=0.7\textwidth, clip=true, trim= 0.1in 0.3in 0.3in 0.5in ]{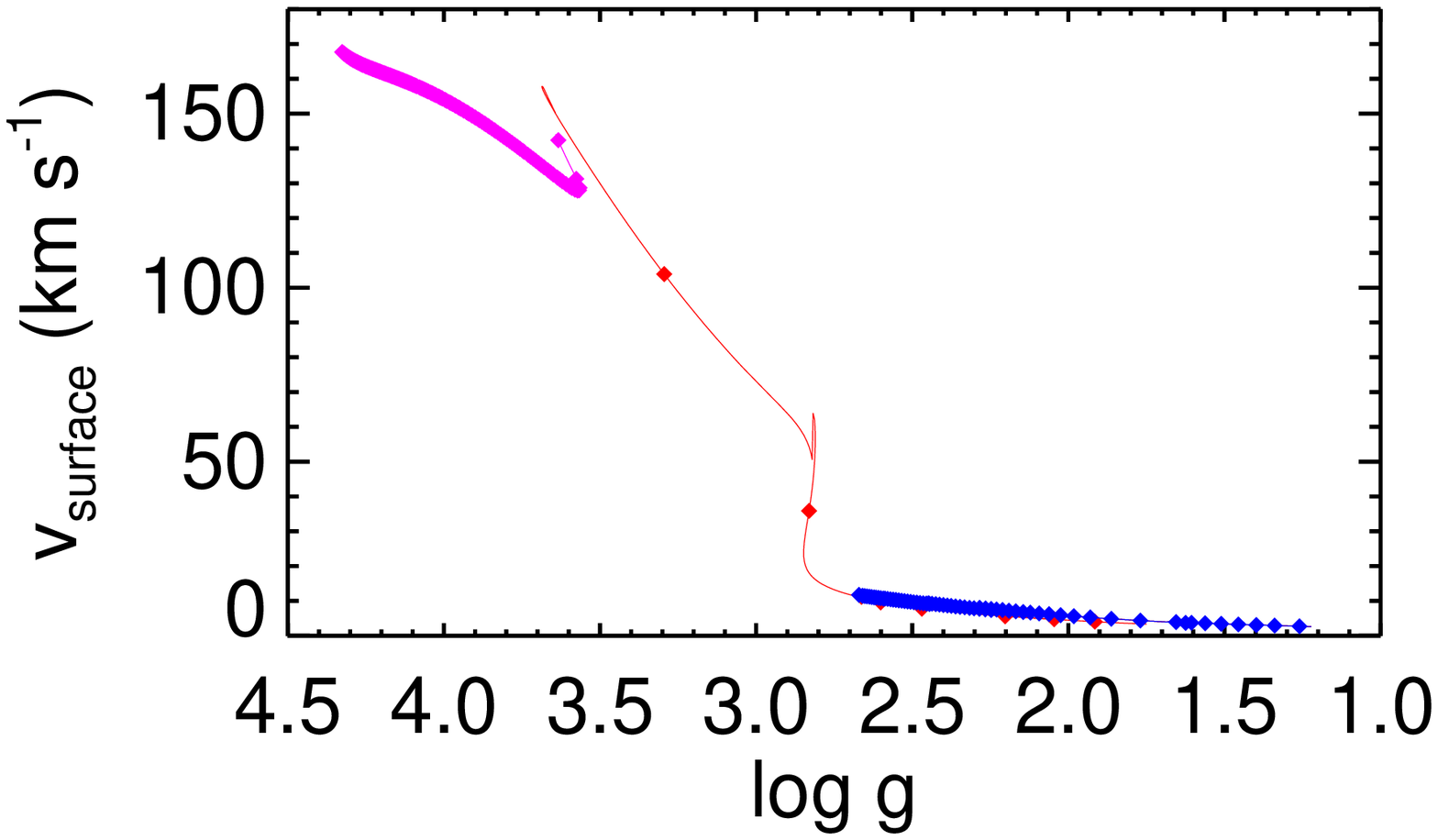}}
\subfigure{\llap{\raisebox{3.5cm}{\includegraphics[height=2.4cm,clip=true, trim=-15.cm 0.5cm -6cm 1.0cm]{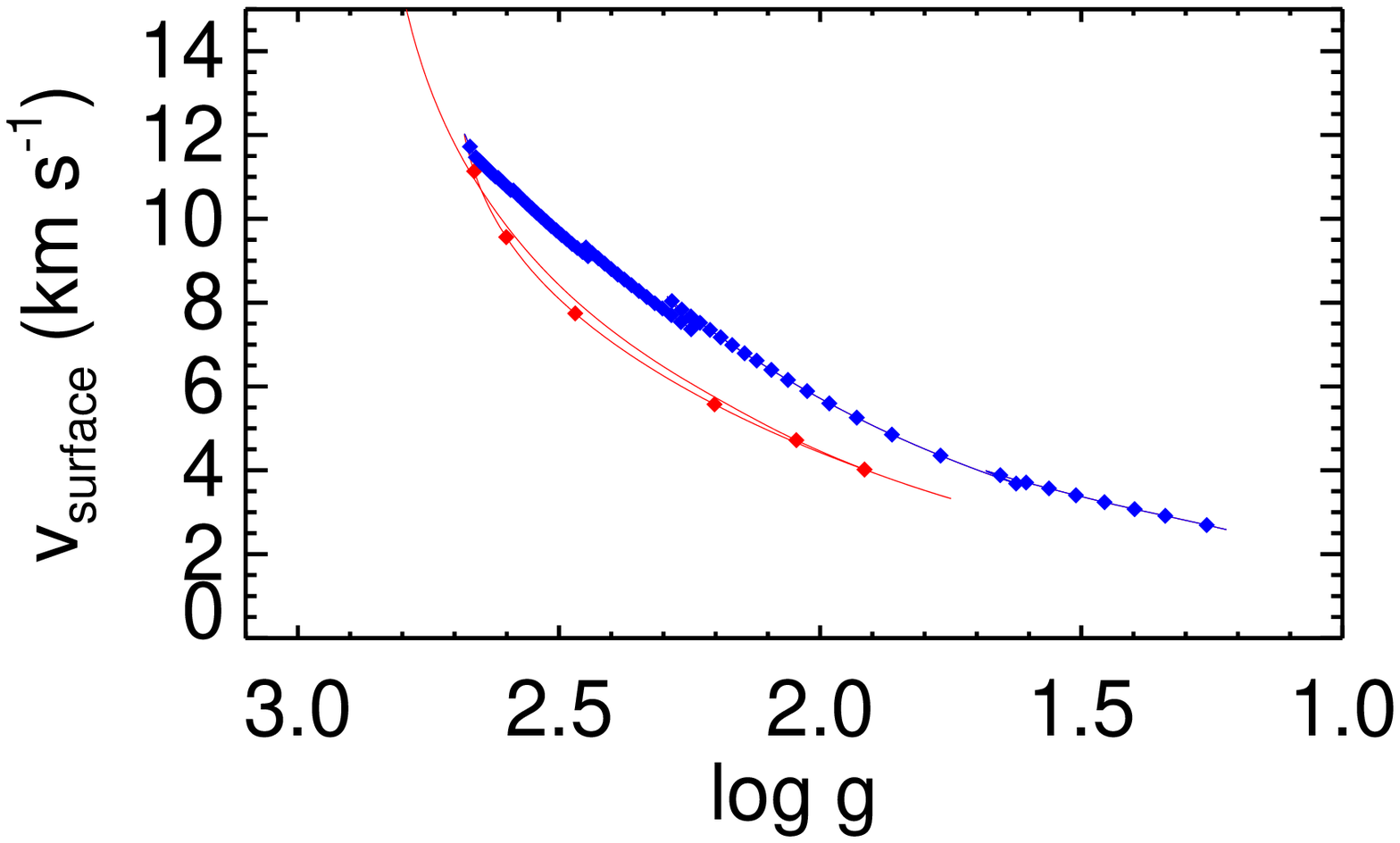}}}}

\subfigure{\includegraphics[width=0.7\textwidth, clip=true, trim= 0.1in 0.3in 0.3in 0.3in ]{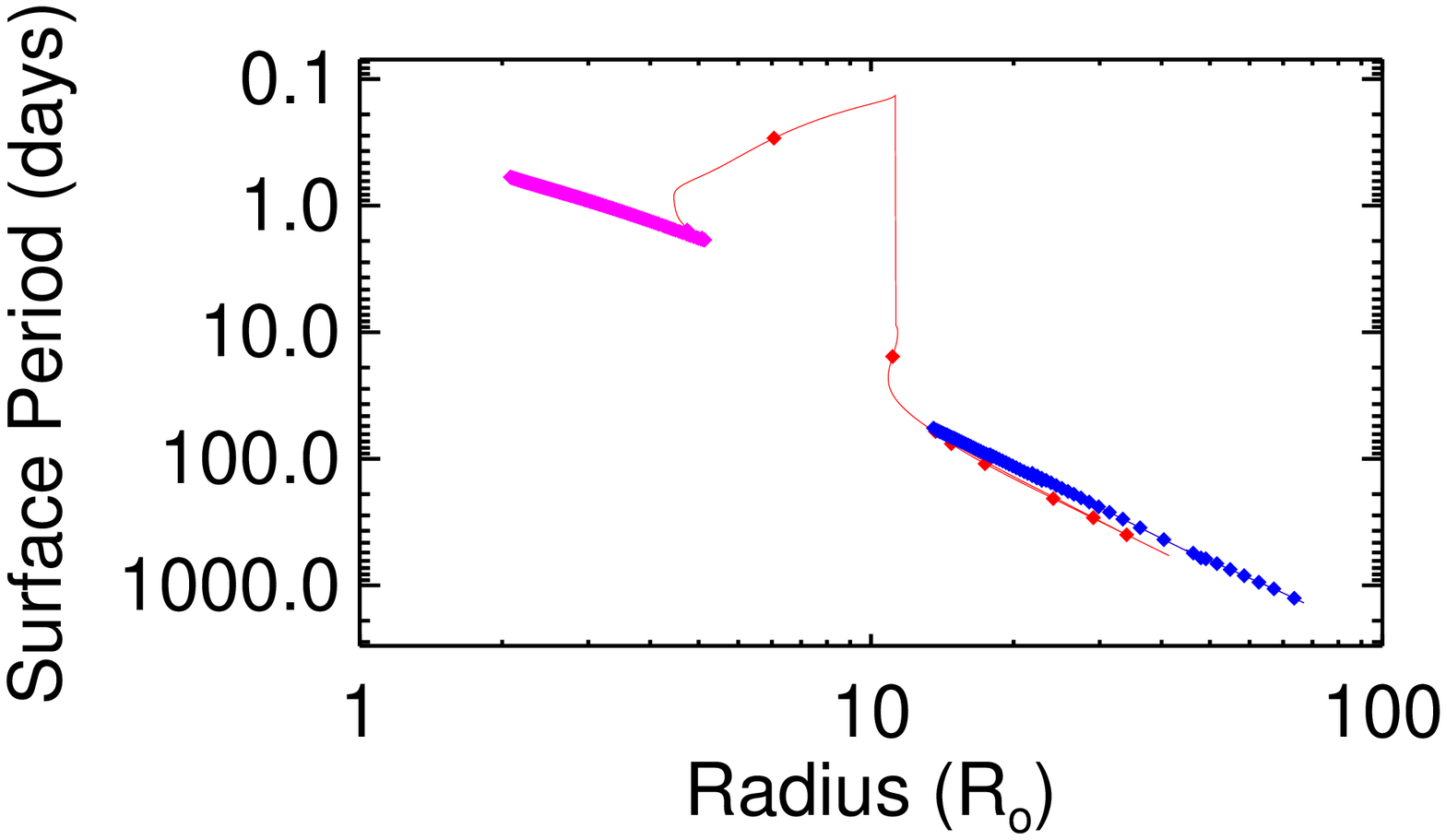}}
\subfigure{\llap{ \raisebox{2cm}{\includegraphics[height=2.4cm,clip=true, trim=-5.cm 0.8cm -25cm 0cm]
             {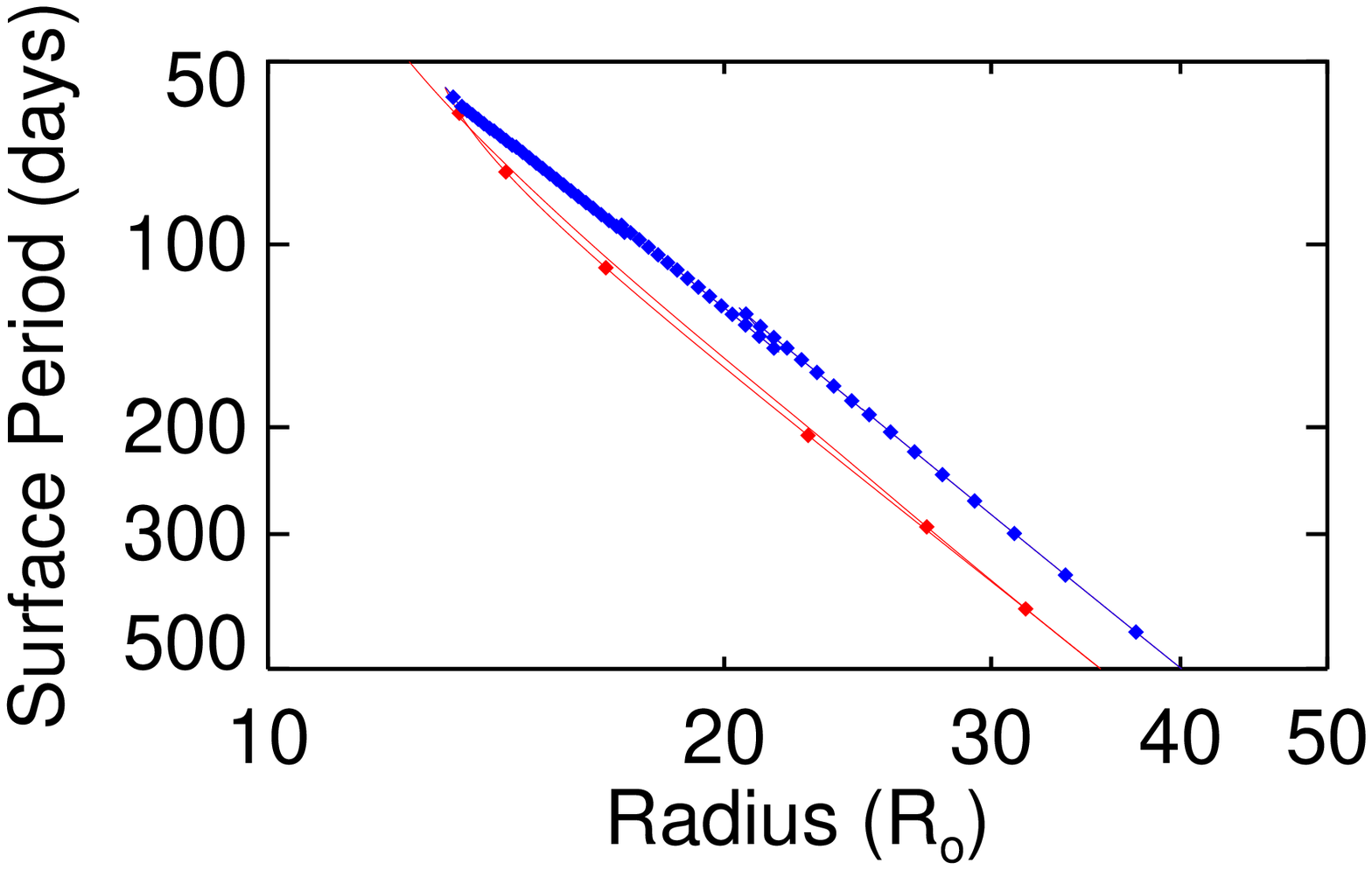}}}}

\caption{The evolution of a 3.0 M$_\sun$ moderate rotator ($\sim$ 150 \kms\ at the end of the main sequence). On all three plots, stars move from the main sequence (pink) to the first ascent giant branch (red). They ascend the giant branch and contract back down to the core helium burning phase (blue). Diamonds indicate steps of 2 Myrs in all cases, and stars generally move from left to right on these plots. The top panel show the star's evolution across the Hertzsprung-Russell diagram, the middle panel shows the evolution of the surface rotation rate, and the bottom shows the evolution of the surface rotation period. Each panel contains an inset which focuses in on the core helium burning phase.}
\label{Fig:Intuition}
\end{minipage}
\end{figure*}

\subsection{Standard Evolution} \label{ssec3:standard}
The evolution of our models on the HR diagram is illustrated in Figure \ref{Fig:Intuition}, along with the corresponding angular momentum evolution. Stars between 2 and 3 M$_\sun$ have a convective core on the main sequence, and their luminosity and radius grows as the mean molecular weight increases. This also causes the rotation rate to slow slightly on the main sequence (pink).  When this core is exhausted, the structure of the star readjusts, which causes a rapid increase in rotation. This transient increase is quickly followed by substantial slowdown as the core of the star contracts, the envelope expands, and the star rapidly crosses the Hertzsprung gap (on roughly the Kelvin-Helmholtz timescale of the core). At this point, hydrogen is being burned in a shell, and the star ascends the giant branch; the surface gravity and temperature decrease as the radius increases during this phase (red). During the red giant phase, the substantial increase in the moment of inertia slows the star down dramatically. Eventually the core grows to a size large enough to ignite helium burning, either degenerately in stars below about 2.25 M$_\sun$ or nondegenerately in stars above that mass. The star then contracts again to about ten solar radii, and starts burning helium in its core (blue). Combining the structural changes and the increase in radius, maximal rotation rates in the core helium burning phase are approximately a factor of ten lower than the end of the main sequence rotation rates assuming there is no angular momentum loss. More evolved core-helium burning stars would rotate even more slowly as they expand further during the core helium burning phase.

\subsection{Thermal Wind}\label{ssec3:thermal}
Even in the absence of a magnetized wind, one would expect angular momentum loss to accompany the post-main-sequence mass loss ($\dot{J}=-\frac{2}{3}\omega R^2 \frac{dM}{dt}$). As discussed in Section \ref{ssec:thermal}, we consider the effect on rotation of a Reimers mass loss and a thermal wind. 
Stars in our mass range experience relatively little mass and angular momentum loss, with mass loss rates of order 
2$\times 10^{-10}$ M$_\sun$ year$^{-1}$ at the zero age horizontal branch, and less than 10$^{-8}$ M$_\sun$ year$^{-1}$ for the most massive stars at the tip of the giant branch. In total, these stars lose less than 0.3\% of their angular momentum content to a thermal wind on the first ascent giant branch, and between 2\% (low mass) and 4\% (high mass)
 in the core helium burning phase.  This difference corresponds to an effectively undetectable change ($<$1 \kms) in the surface rotation rate. Given this small impact, we therefore do not include this loss source for our models, especially in light of the larger effects from magnetized winds discussed in the next section.

\begin{figure}[t]
 \centering
\includegraphics[width=0.5\textwidth, clip=true, trim= 0.5in 0.2in 0.1in 0.2in ]{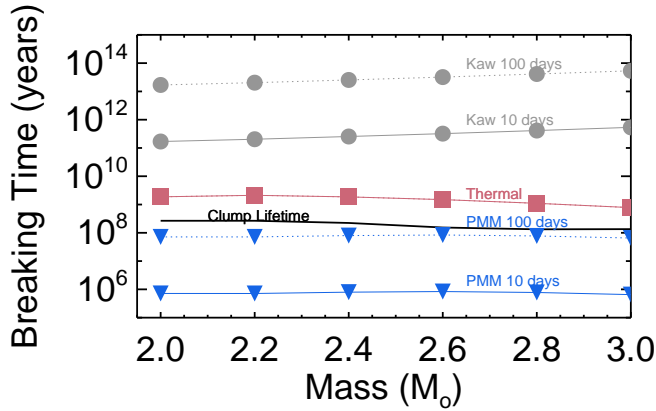}
\caption{The breaking timescale computed as the total angular momentum content of the star divided by the angular momentum loss rate. Dotted lines assume a rotation period of 100 days, solid lines indicate periods of 10 days. We show the lifetime of the core helium burning phase as a function of mass for comparison. While the breaking time of the PMM loss law is shorter than the length of this phase, this is not true of the thermal or Kawaler wind loss laws, which do not substantively break these stars.}
\label{Fig:Jdot}
\end{figure}

\subsection{Magnetized Winds}\label{ssec3:magnetized}
A magnetized wind can enforce co-rotation to the Alfv{\'e}n radius, resulting in a large amplification in the angular momentum loss rate per unit mass.  Angular momentum loss rates from such winds are typically scaled relative to the sun, with parametrizations of the effective mean surface magnetic field strength and mass loss rates in terms of the rotation rate and structural variables. Recovering the \citet{Skumanich1972} spin down rate for solar analogs requires a torque that scales as $\omega^3$, the scaling predicted from the solar wind by \citet{WeberDavis1967}. The \citet{Kawaler1988} solution predicted a spindown rate that was only a weak function of radius, a conclusion challenged by \citet{ReinersMohanty2012} on the grounds that an alternative scaling of magnetic field strength was preferred. \citet{Matt2012} derived a more rigorous solution for the topology of the solar wind, also resulting in a stronger predicted relative dependence on stellar mass and radius.  Finally, the formulations by \citet{vanSadersPinsonneault2013} and \citet{GalletBouvier2013} also imply loss rates which are sensitive to the convective overturn timescale. We therefore have two general classes of spindown models: the \citet{Kawaler1988} family, with loss rates that depend only weakly on the global stellar parameters, and the more modern loss prescriptions, which predict large differences in rates between solar analogs and evolved stars.

We show in Figure \ref{Fig:Jdot} an estimate of the time for magnetic breaking to occur, computed by taking the total angular momentum content of the star and dividing by the loss rate at the Zero Age Horizontal Branch. For the Kawaler case, the breaking time is substantially longer than the lifetime in the core helium burning phase, and we find that this prescription has a $<$1\% effect on the total angular momentum content of our secondary clump stars, even in the most rapidly rotating rigid case. 

We also consider the PMM wind loss law, which has a different structural dependence. In particular, the strong dependence on radius ($(\frac{R}{R_\sun})^{3.1}\sim13^{3.1}\sim2\times10^3)$, and explicit dependence on convective overturn timescale ($(\frac{\tau}{\tau_\sun})^{2}\sim5^2\sim25$) and luminosity ($(\frac{L}{L_\sun})^{0.56}\sim100^{0.56}\sim13$) serve to increase the magnetic breaking in our core helium burning stars by almost six orders of magnitude. This produces enough loss that the breaking time becomes comparable to the lifetime in the core helium burning phase, and significant angular momentum loss can happen on the secondary clump.

\begin{figure}[t]
 \centering
\includegraphics[width=0.5\textwidth, clip=true, trim= 0.5in 0.2in 0.1in 0.5in ]{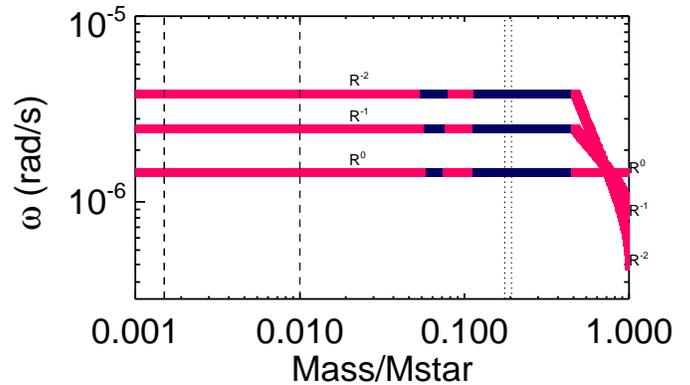}
\caption{The rotation profile of a 2.4 M$_\sun$ star with a rotation velocity of 150 \kms\ at the end of the main sequence, shown in the core helium burning phase at log(g)=2.75 for three different convection zone rotation profiles $(\omega \propto R^0, \omega \propto R^{-1}, \omega \propto R^{-2} )$. Pink regions are convective, blue regions are radiative. Thick black vertical dashed lines mark the approximate extent of the core region where seismology is most sensitive. Thin vertical dotted lines indicate the range of the hydrogen burning shell. Note that an increase in the amount of differential rotation in the convective zone causes a corresponding increase in the rotation rate of the core at fixed angular momentum content.}

\label{Fig:Profiles}
\end{figure}

\subsection{Differential Rotation}\label{ssec3:diffrot}
In evolved stars, the overwhelming majority of the moment of inertia is in the convective envelope, especially for lower mass and lower gravity stars. This means that the choice of the rotation profile in the surface convection zone has a much larger effect on the angular momentum content of the star than the choice of the radial rotation profile in the core. Specifically, we find that the allowing local conservation of angular momentum in the core changes the predicted surface rotation rate by less than 10 percent (2 \kms) in the most extreme case. In this work, we therefore only consider the possibility that the surface convection zones of these stars rotate differentially, with the rotation rate going as some power of the radius (see Figure \ref{Fig:Profiles}). At fixed angular momentum content, stronger differential rotation in the envelope yields a slower surface rotation rate and a faster core rotation rate. Additionally, in the case where both angular momentum loss and radial differential rotation are happening simultaneously, the slower surface rotation rate will reduce the angular momentum loss rate, causing the star to retain more total angular momentum and possibly even rotate faster than it would have if only loss had been acting.

\begin{figure*}[t]
\begin{minipage}{\textwidth}
 \centering
\subfigure{\includegraphics[width=.49\textwidth, clip=true, trim= 0.7in 0.2in 0.1in 0.1in ]{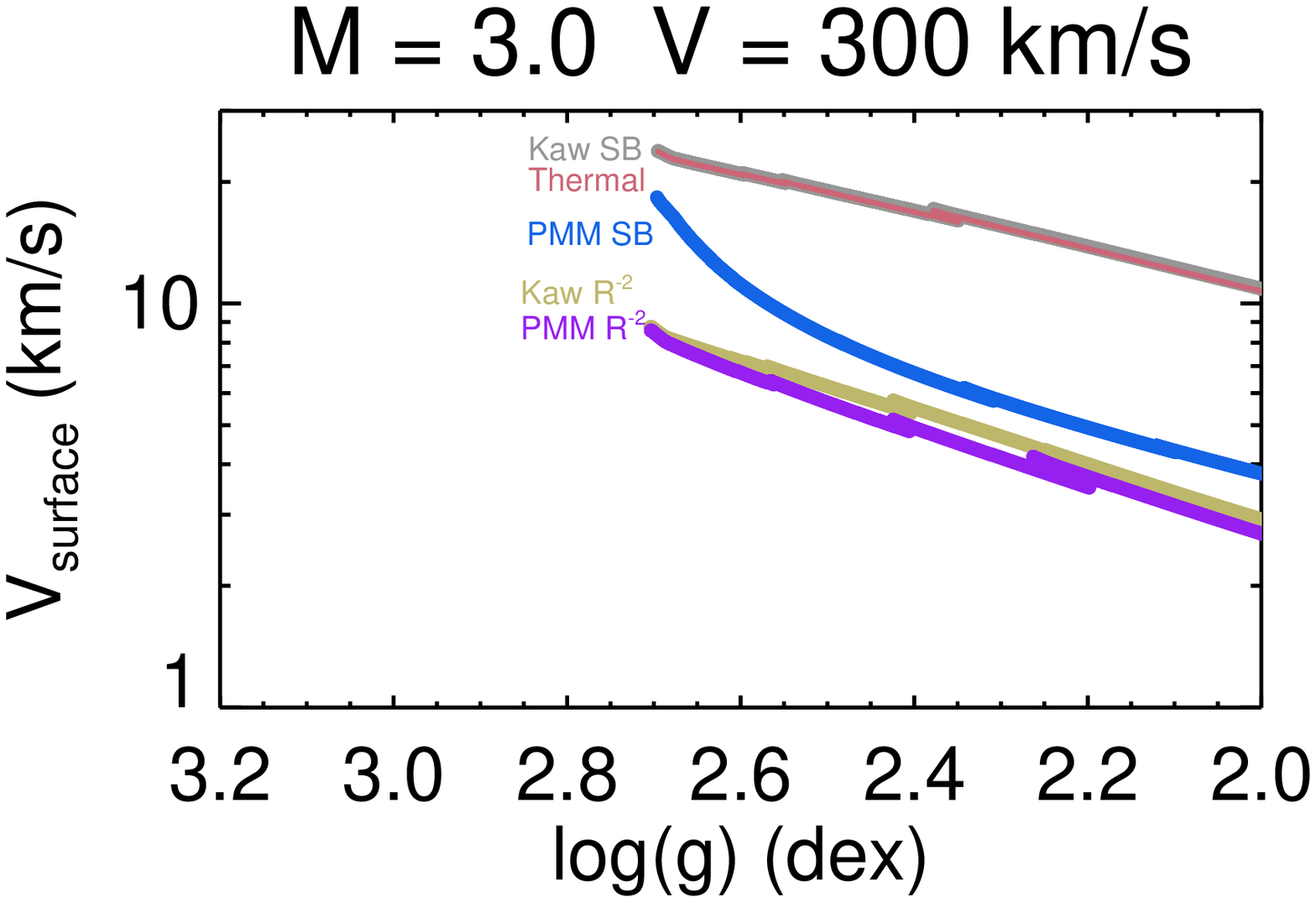}}
\subfigure{\includegraphics[width=.49\textwidth, clip=true, trim= 0.7in 0.2in 0.1in 0.1in]{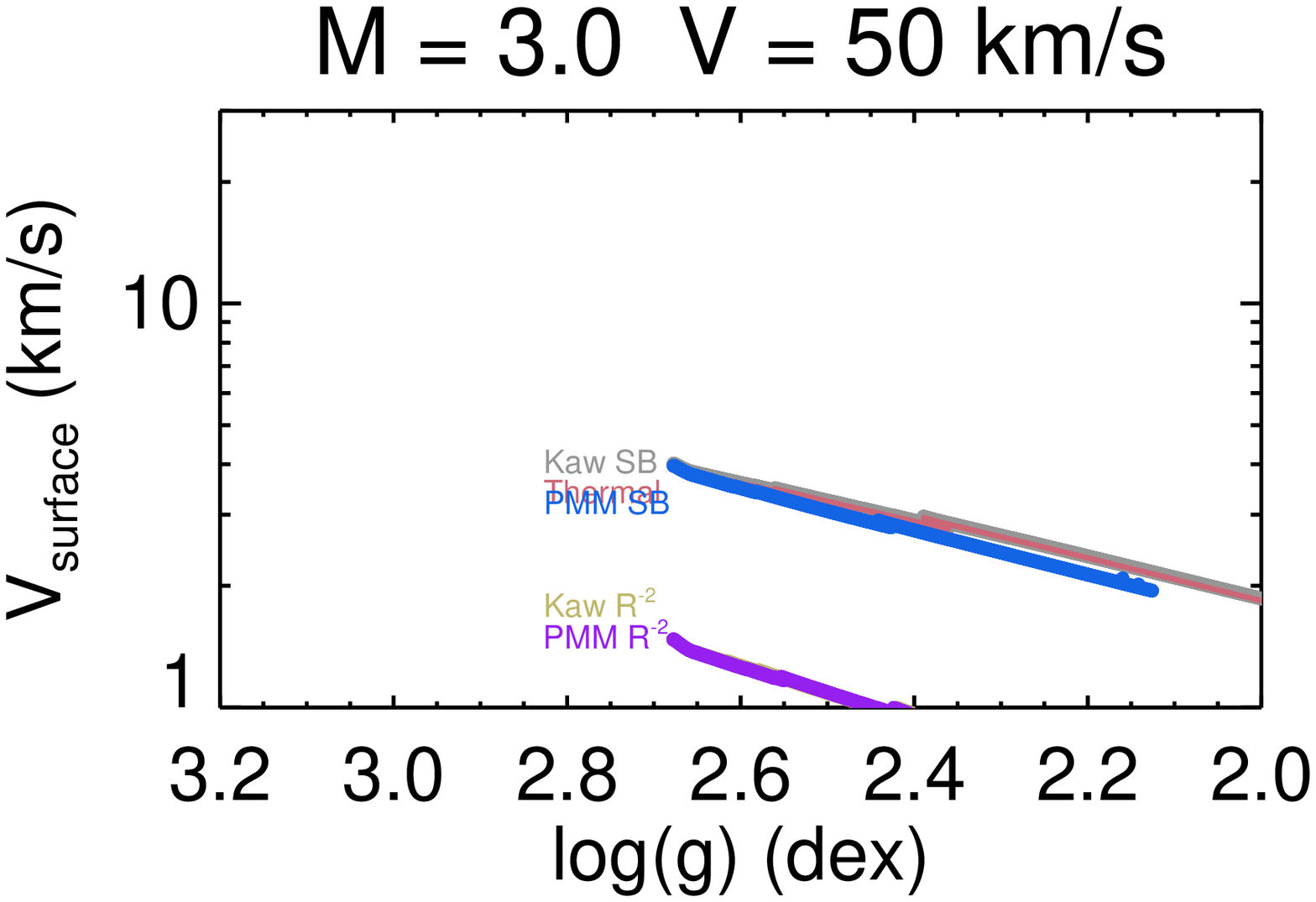}}
\subfigure{\includegraphics[width=.49\textwidth, clip=true, trim= 0.7in 0.2in 0.1in 0.1in]{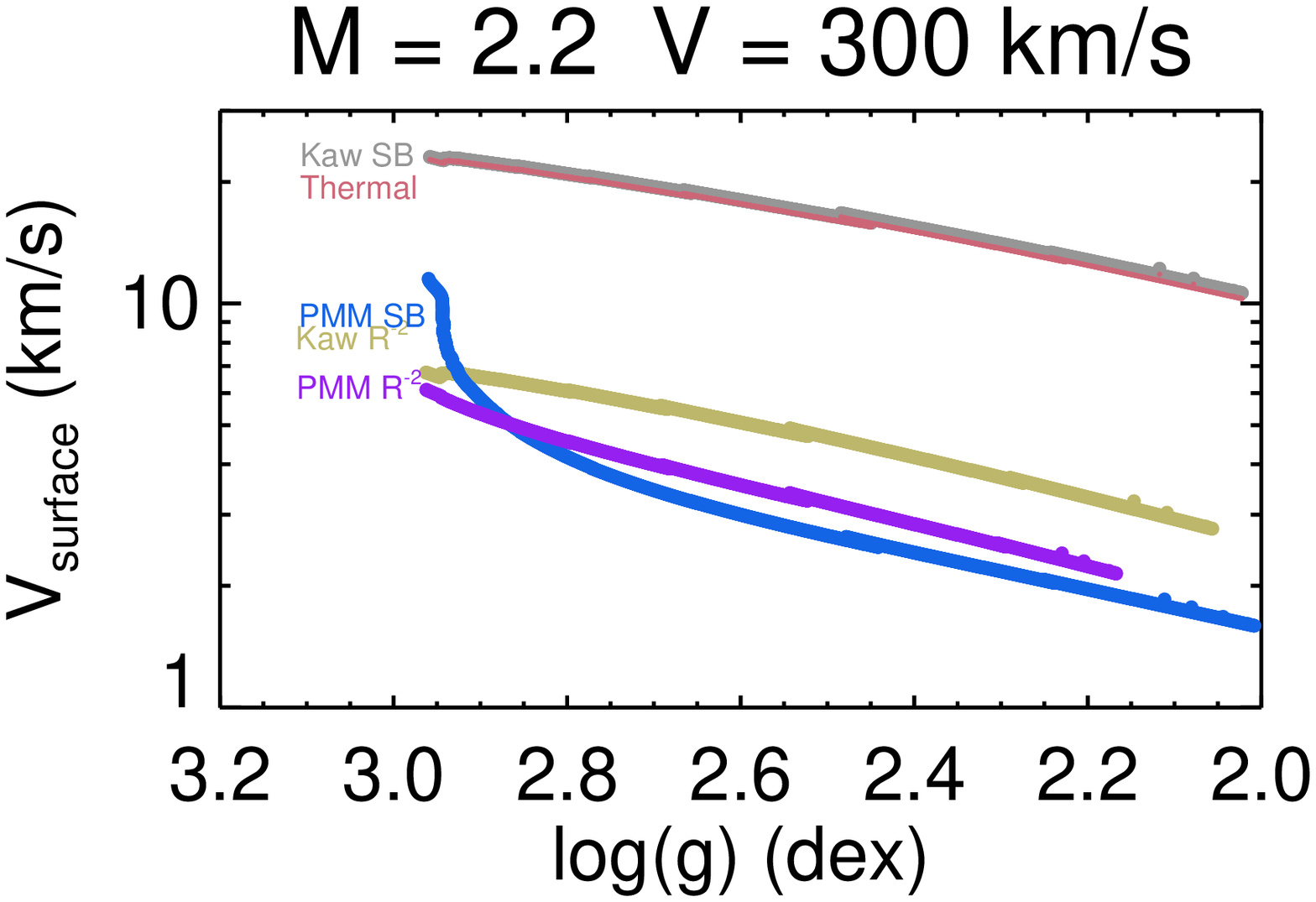}}
\subfigure{\includegraphics[width=.49\textwidth, clip=true, trim= 0.7in 0.2in 0.1in 0.1in]{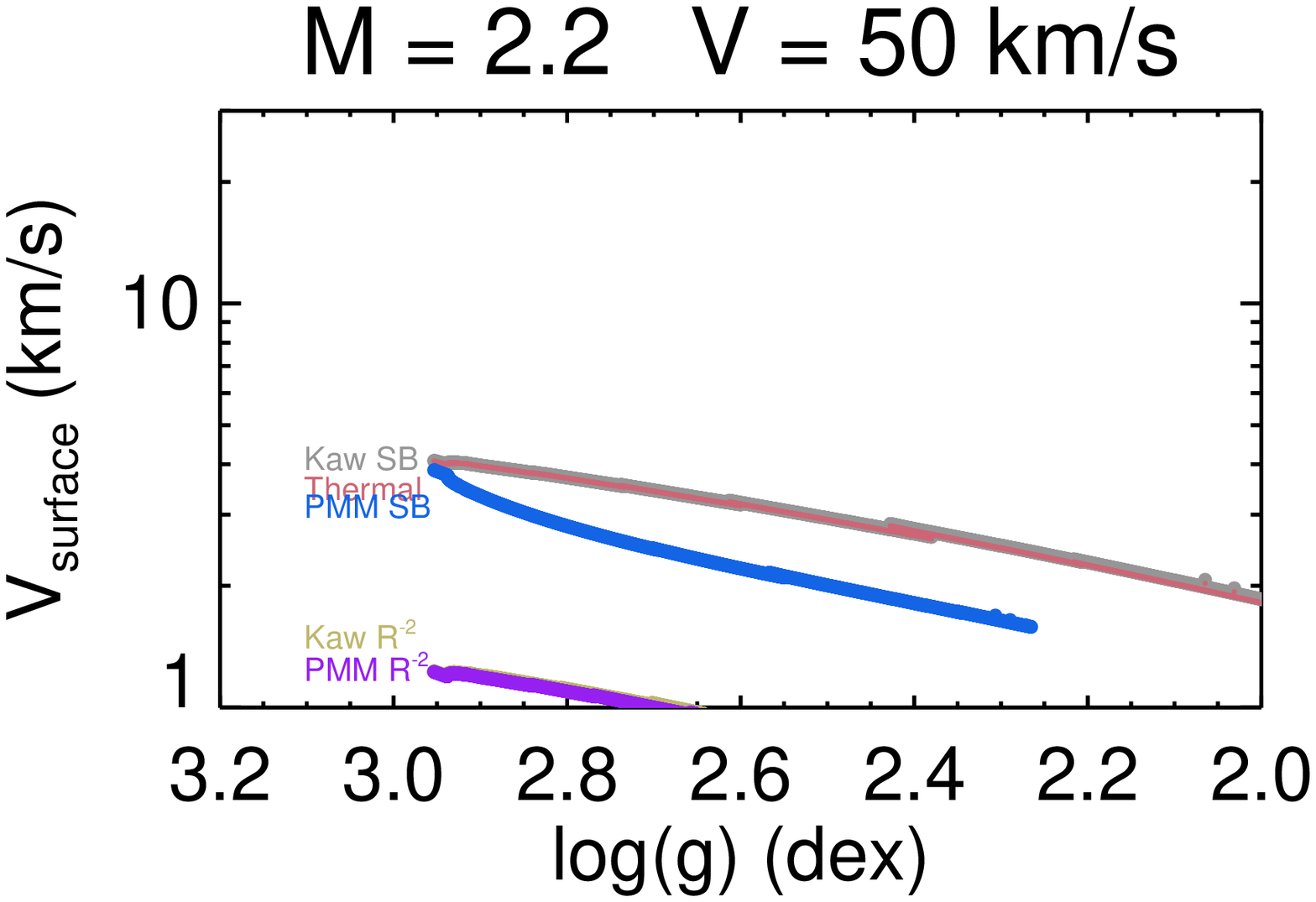}}

\caption{Rotation rate for each of our physics cases as a function of gravity in the core helium burning phase. We show both the fast (300 \kms, left) and slow (50 \kms, right) starting conditions for 2.2 M$_\sun$ (bottom) and 3.0 M$_\sun$ (top) cases. }
\label{Fig:Loss}
\end{minipage}
\end{figure*}

\subsection{Observed Trends}\label{ssec3:observed}

For each of the rotation and loss cases discussed in Section \ref{sec:RotPhysics} (see Table \ref{Table:rotation}), we show the predicted rotational evolution in the core helium burning phase as stars evolve to lower surface gravity in Figure \ref{Fig:Loss}. In general, stars slow down as gravity decreases because the expansion of the radius of the star causes a corresponding increase in the moment of inertia. On top of this, the existence of strong angular momentum loss predicts additional slow down of the surface rotation during the core helium burning phase, steepening the dependence of the predicted rotation rate on gravity. This effect is strongest in fast rotating and lower mass (longer-lived) stars. Differential rotation, in contrast, slows down the absolute rotation rate, but does not substantially change the dependence of the loss on gravity.

\begin{figure}[h]
 \centering
\subfigure{\includegraphics[width=0.49\textwidth, clip=true, trim= 0.5in 0.2in 0.1in 0.0in ]{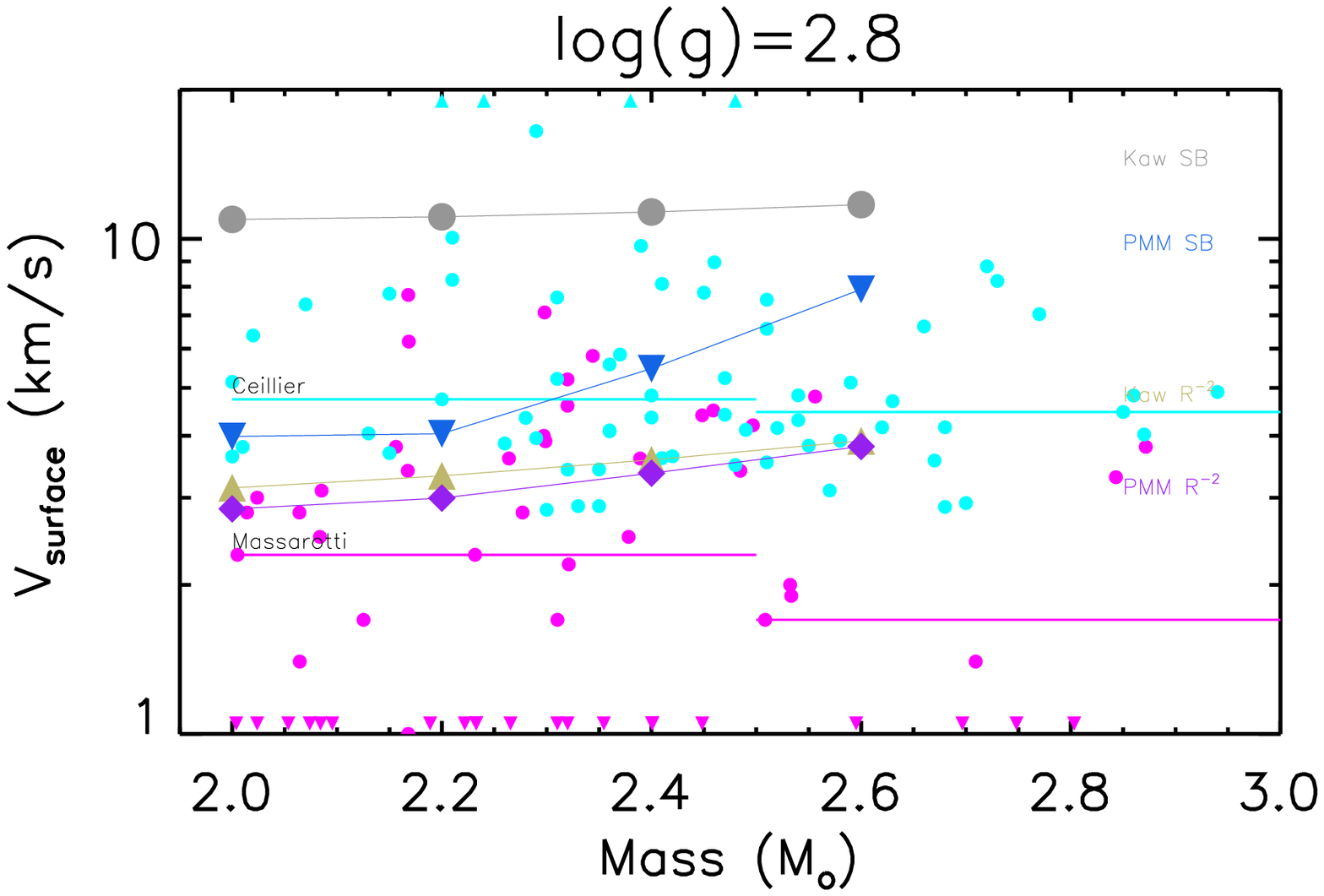}}
\subfigure{\includegraphics[width=0.49\textwidth, clip=true, trim= 0.5in 0.2in 0.1in 0.0in ]{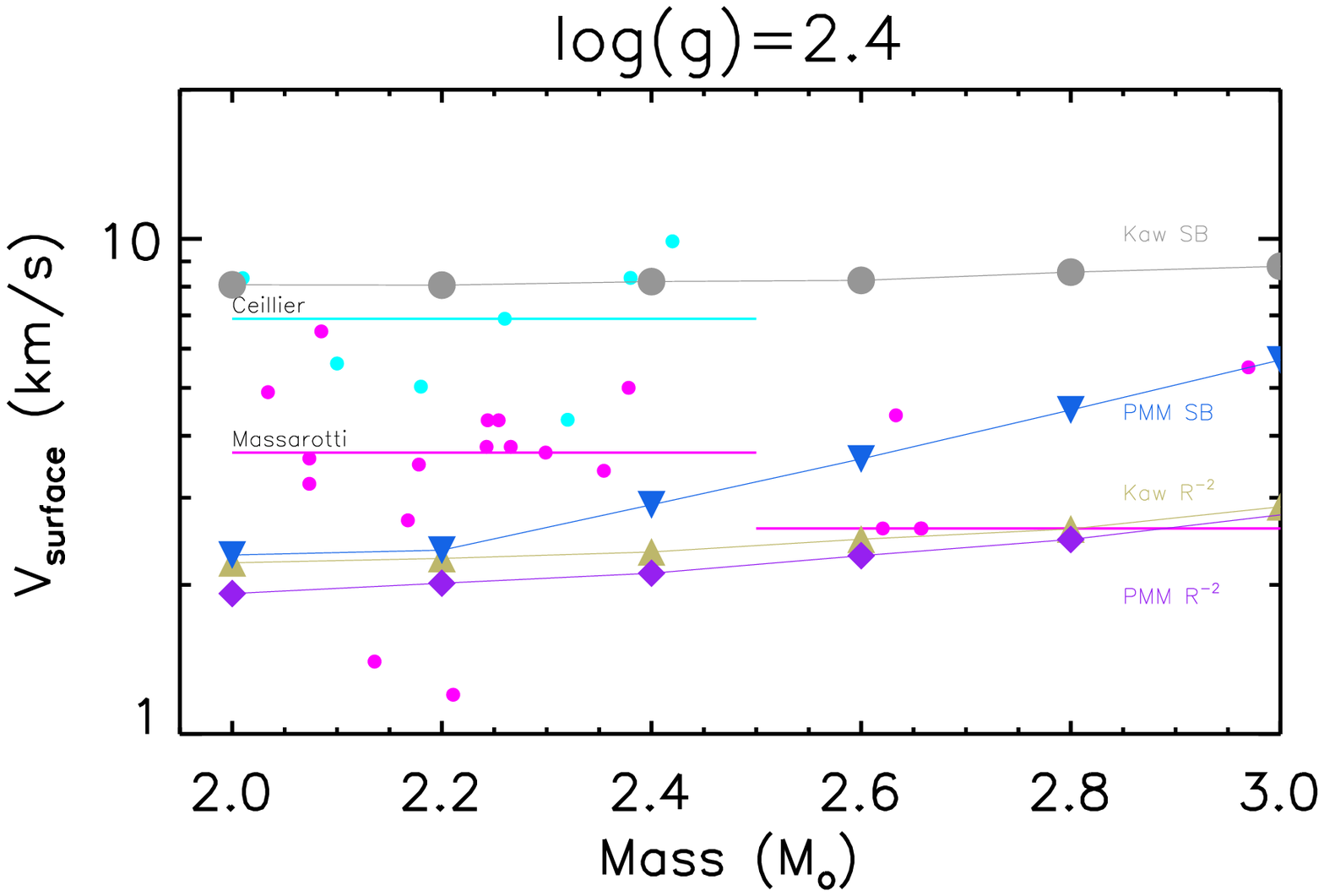}}
\caption{{The predicted surface rotation of a median rotator (150 \kms) as a function of mass for each of our theoretical models. We show the predictions at a log(g) of 2.8 dex (top) and 2.4 dex (bottom). We also compare to the measured median rotation rates of each sample divided into a low (2.0-2.5 M$_\sun$) and high(2.5-3.0 M$_\sun$) mass bin in the appropriate gravity ranges (log(g) between 3.0 and 2.6 in the top plot and between 2.6 and 2.2 in the bottom plot).}} 
\label{Fig:mass}
\end{figure}

Over the mass range we consider, the initial distribution of rotation rates is not strongly mass dependent. In the absence of substantial loss or radial differential rotation, we therefore expect minimal trends with mass in surface rotation at fixed surface gravity (Figure \ref{Fig:mass}). However, in the case of significant angular momentum loss, the factor of a few difference in lifetime begins to imprint a trend in surface rotation rate with stellar mass, with lower mass (longer lived) stars rotating more slowly than more massive, shorter lived stars. If radial differential rotation in the surface convection zone is allowed, the overall rotation rate is reduced at fixed mass and gravity, and the effects of angular momentum loss are also reduced, substantially decreasing any mass trend. We also show in Figure \ref{Fig:mass} the observed {rotation velocities} 
of two of our samples in bins of mass and surface gravity. We see no strong mass trends in the observed samples. However, we caution that measuring the true average rotation rate can become extremely difficult at slow rotation rates since other sources of spectral line broadening  become important (v sin (i) measurements) and controlling for long term systematics in light curves becomes more difficult (period measurements). {Additionally, given the relatively small number of stars in the sample, we have used large bins in surface gravity, and have not included the uncertainties on the masses or surface gravities of the stars in this comparison, which could also help to disguise trends in the real underlying population. We therefore consider this comparison suggestive, rather than definitive, in our attempt to understand the underlying physics of rotation in this regime.}

Because we do not know the dependence of the surface rotation rates on the main sequence as a function of metallicity, it is impossible to construct a true prediction of the dependence of rotation on the secondary clump as a function of metallicity. We note that structural and lifetime differences for models between [Fe/H]=-0.3 and [Fe/H]=+0.3, a reasonable range for the young stars in this mass range, do not substantially alter the predicted dependence of rotation on mass and surface gravity.

\section{Predicted Populations} \label{sec:Distributions}

It has long been known from work on clusters that giants in our mass range rotate slower than simple models with solid-body rotation and solar-like loss would predict \citep{GrayEndal1982, Carlberg2016}.
However, our new analysis is performed on data sets that are orders of magnitude larger and span a wider range of masses and surface gravities. Additionally, we can also explore the predictions of physically motivated models. 
These tests require the construction of distributions of initial rotation rates as a function of mass (and ideally metallicity), followed by a comparison of these predictions to the measured core and surface rotation rates in evolved stars.  Unfortunately, as discussed in Section \ref{Data}, all of our samples are less than ideal, with difficult-to-quantify selection effects. We therefore build a general predicted distribution, neglecting sample selection biases, and then check to see whether the global properties (such as distribution on the HR diagram) are close to that of the full sample; we will demonstrate that the two are at least broadly consistent. We return to the question of the impact of sample selection bias in the discussion section.

\begin{figure}[h]

 \centering
\subfigure{\includegraphics[width=0.44\textwidth, clip=true, trim= 0in 0.2in 1.3in 1.7in ]{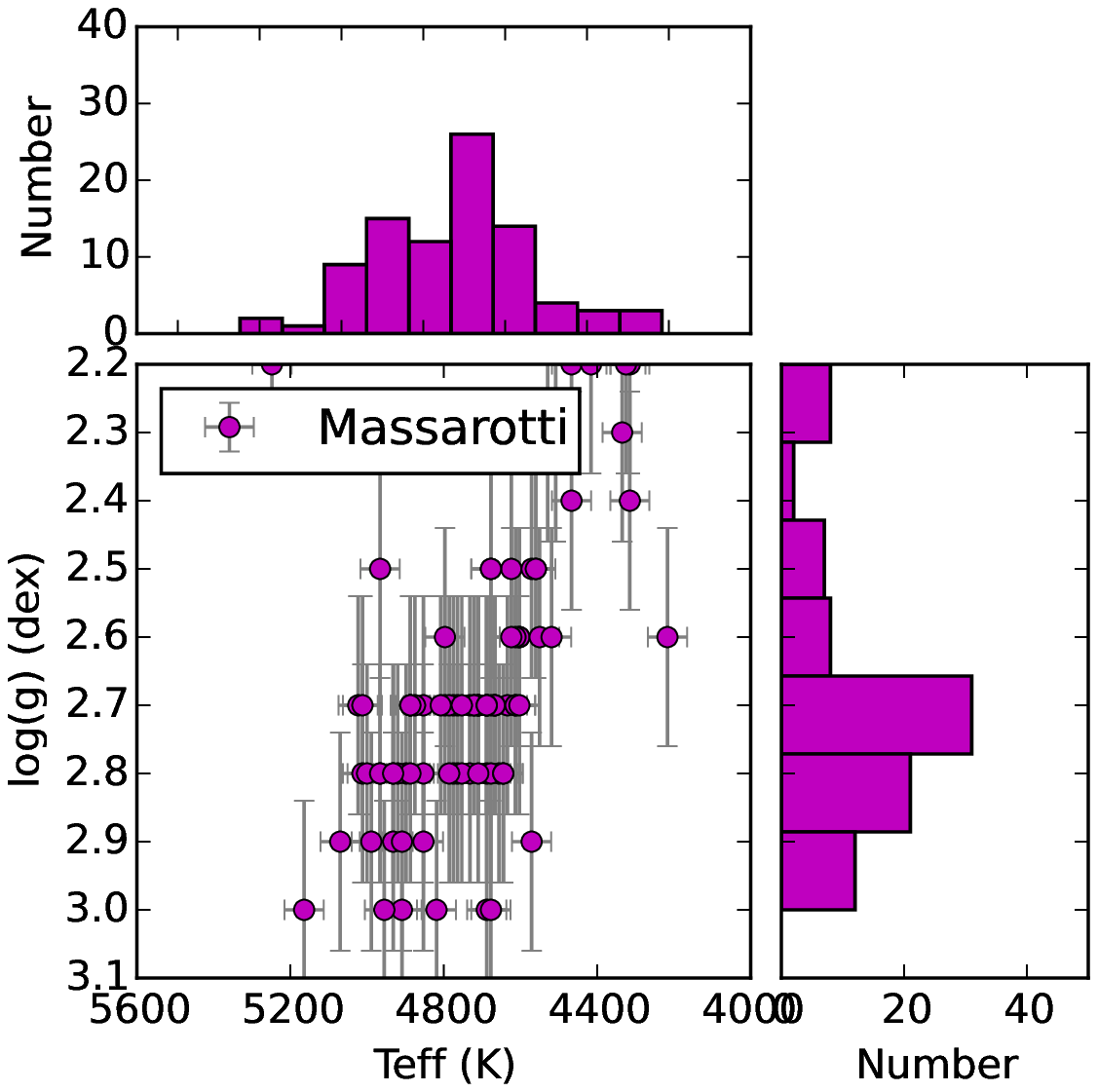}}
\subfigure{\includegraphics[width=0.44\textwidth, clip=true, trim= 0in 0.2in 1.3in 1.7in ]{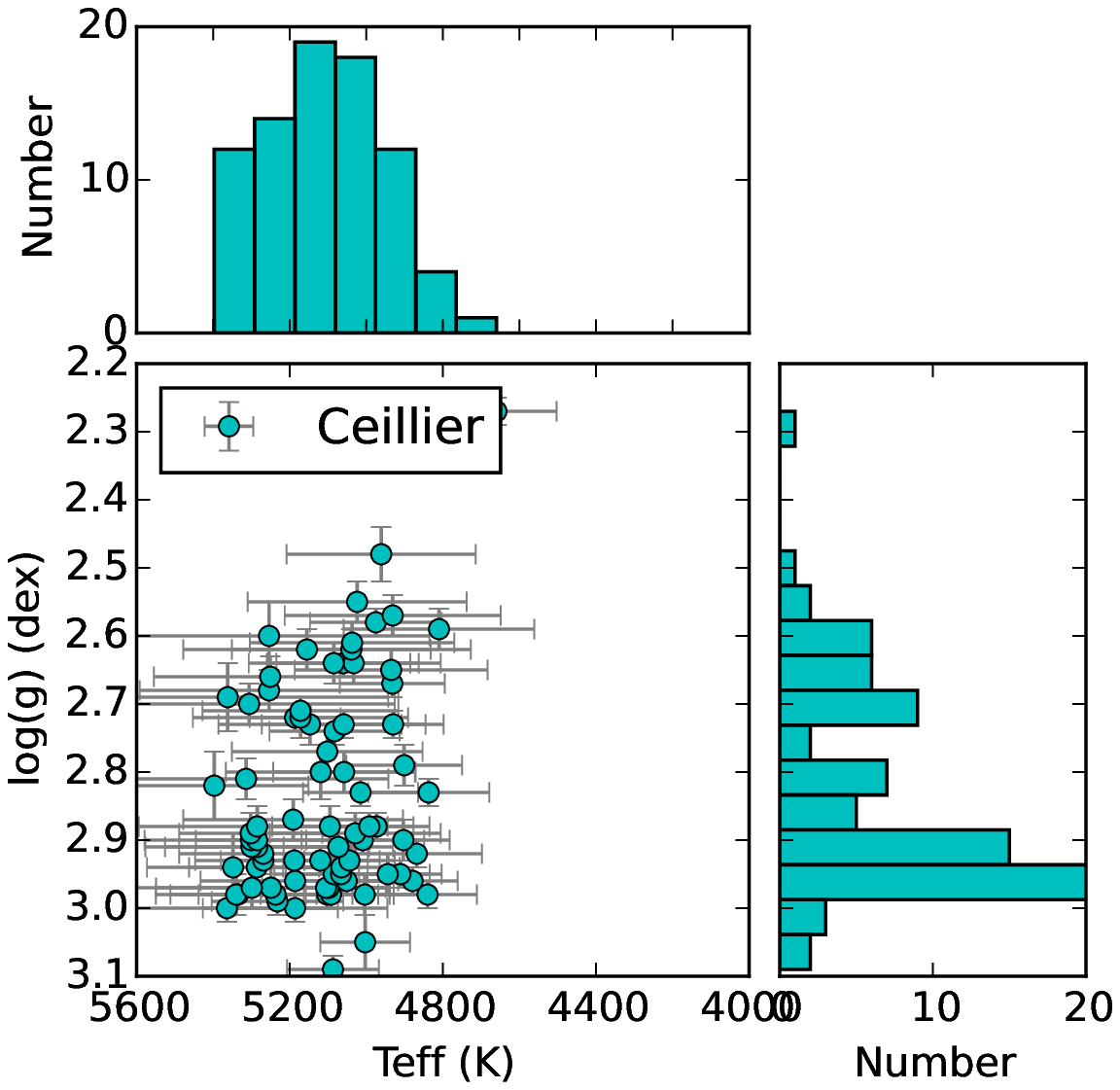}}
\subfigure{\includegraphics[width=0.44\textwidth, clip=true, trim= 0in 0.2in 1.3in 1.7in ]{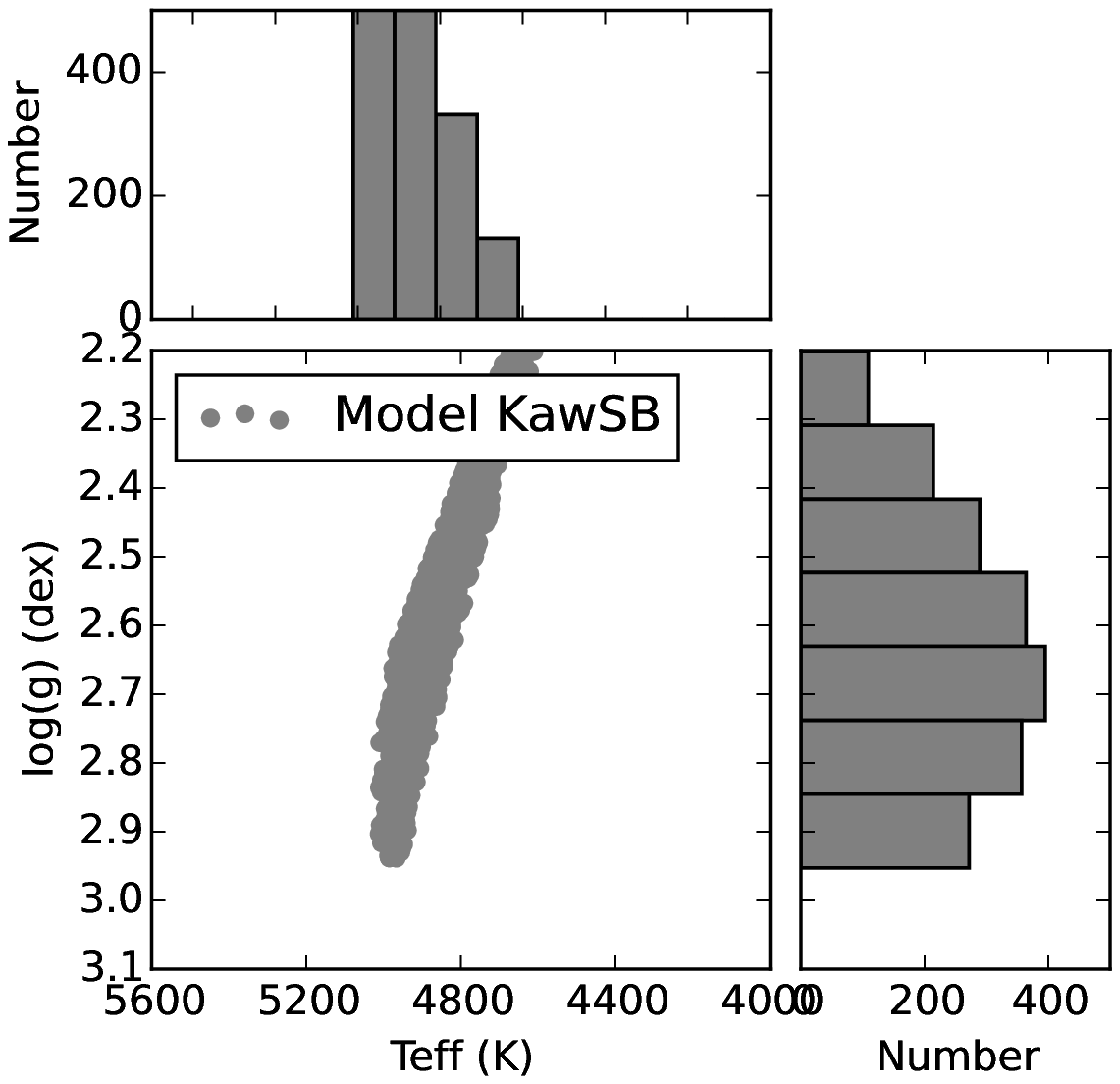}}
\caption{HR diagram position of the stars in the \citet{Massarotti2008} (top), and \citet{Ceillier2017} (middle) samples compared to the location of the stars in our randomly drawn distribution (bottom). }
\label{Fig:HRdiagram}
\end{figure}

To construct our predicted distributions of rotation rates, we forward model rotation as a function of mass, initial rotation rate, and age for different angular momentum evolution scenarios.  In particular, we constructed grids of models run for stars between 2.0 and 3.0 M$_\sun$ (in increments of 0.2 M$_\sun$) and end of the main sequence rotation velocities from $<$1 \kms\ to 300 \kms\ (in increments of 50 \kms). Each of these cases is then used to map a given initial condition onto a predicted surface rotation rate in the secondary clump as a function of mass, initial rotation and surface gravity for each scenario. We model the rotation rate distribution by taking (2000) random draws from the double Maxwellian of end of the main sequence rotation rates inferred for stars between 2.3 and 2.7 solar masses by \citet{ZorecRoyer2012}. We infer the mass distribution using the same number of draws from a Salpeter IMF with limiting masses of 2.0 and 3.0 M$_\sun$. We then generate the surface gravity distribution by drawing from a flat age prior between the age of the smallest radius after the tip of the giant branch (beginning of the clump) and the age when log(g) equals 2.2 (our definition for the end of the clump).
Finally, we draw a random inclination angle for each star from a flat prior in cosine(i) in order to compare to the measured v sin(i) distributions. 
At this point, we have defined a mass, surface gravity, main sequence rotation rate, and inclination angle for each of our stars.  We then use linear interpolation within our grid to find the core helium burning rotation rate for each of our simulated stars and plot the expected distribution of rotation rates. We show in Figure \ref{Fig:HRdiagram} the locations of stars in our measured and predicted distributions.

\subsection{Comparison of Surface Distributions} \label{DistCompare}
\begin{figure}[h]
 \centering
\includegraphics[width=0.49\textwidth]{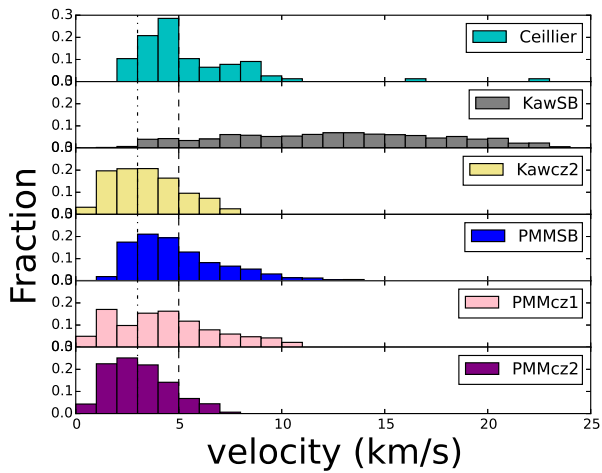}
\caption{{Comparison of the measured surface velocities from \citet{Ceillier2017} to the predicted distributions for each of our cases. Models with Kawaler wind loss and rigid rotation clearly rotate faster than the observations, but measurement uncertainties and sample selection effects make it impossible to distinguish between the other options.}}
\label{Fig:Ceillier}
\end{figure}

\begin{figure}[h]
 \centering
\includegraphics[width=0.49\textwidth]{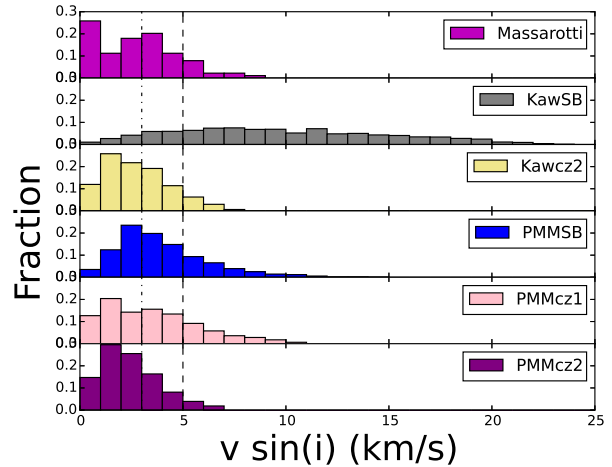}
\caption{{Comparison of the measured surface velocities from \citet{Massarotti2008} to the predicted distributions of a field population for each of our cases. As in Figure \ref{Fig:Ceillier}, the Kawaler solid body predictions are clearly faster than the observations, but it is impossible to conclusively distinguish between the models with PMM angular momentum loss, radial differential rotation, and a combination of the two.}}
\label{Fig:Massarotti}
\end{figure}

We show the predicted distributions of surface rotation velocities for each of our {rotation} cases compared to the measured distribution of \citet{Ceillier2017} in Figure \ref{Fig:Ceillier}. The selection effects in this sample are likely to make it hard to measure slow rotation rates (periods greater than 100 days, a rotation velocity less than about 5 \kms\ at these radii), and to enhance the detection of very rapidly rotating stars.
 We show a similar plot in Figure \ref{Fig:Massarotti} using the v sin(i) distributions compared to the \citet{Massarotti2008} data. The \citet{Massarotti2008} sample was selected using \textit{Hipparcos} parallaxes and therefore should not be biased towards either slow or rapid rotators, but has low precision inferred masses; scattering of lower mass first ascent giants into the sample could therefore produce a bias towards slow rotation. The measured distribution is more skewed towards slow rotation than the \citet{Ceillier2017} data, but the correspondence between the different models and the data is similar, {suggesting that sample selection effects are not dramatically affecting our conclusions. We note that our results would also not be substantively changed}  
by the uncertainties of order 10 percent on the initial surface rotation rates of stars in this mass range (see Section \ref{ssec:ZR2012}).

{Our first major conclusion is that the distribution of predicted rotation rates for our base case, a Kawaler wind loss law with rigid rotation (Kawaler SB), are systematically much faster than the observed distributions. We present four other cases: Kawaler cz2, strong convection zone differential rotation and weak loss; PMM SB, rigid convection zone rotation and enhanced loss; and two cases (PMM cz1 and PMM cz2) with enhanced loss and moderate ($\simeq R^{-1}$ or strong ($\simeq R^{-2}$) convection zone differential rotation with depth.All of these models predict surface rotation rates much closer to the measurements, and it is difficult to discriminate between them using only the surface rotation distributions. There are therefore three possible families of solutions: either stars experience significant angular momentum loss (the PMM SB model); they have strong differential rotation with depth, but little loss (the Kawaler SB model); or both (the PMM cz1 and PMM cz2 models).
}

\subsection{Comparison of Core Distributions} \label{DistCompareCore}
\begin{figure}[h]
 \centering
\includegraphics[width=0.49\textwidth]{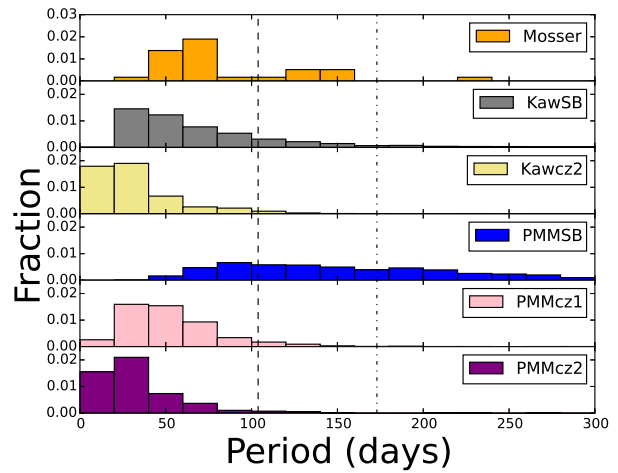}
\caption{{Comparison of the measured core rotation periods \citep{Mosser2012b} to the predicted distributions of a field population for each of our cases, assuming that the core rotates at the same rate as the base of the convection zone. We note that it is possible to speed up the core (shift the distribution to the left) with radial differential rotation in the radiative zone. We find that the maximally rotating cases predict cores rotating faster than observations, and that PMM angular momentum loss predicts cores rotating too slowly, suggesting that moderate differential rotation might fit better than either extreme.}}
\label{Fig:Mosser}
\end{figure}

Core rotation rates {place powerful constraints on the allowed families of rotation models. Our logic proceeds as follows: under the reasonable assumption of a monotonic rotation curve, models with differential rotation in the convection zone are required to have a higher core than surface rate, regardless of the amount of core-envelope decoupling. We can therefore treat the rotation rates at the base of the convection zone as strict lower limits to the measured core rotation rates. }
{We show in Figure \ref{Fig:Mosser} the predicted distribution of core rotation rates for each of these scenarios assuming that the core rotates at the same rate as the base of the convective envelope, compared to the core rotation distribution measured by \citet{Mosser2012b}. Both of the models with minimal angular momentum loss (KawSB and Kawcz2) predict core rotation rates that are faster than the observations. We contend that this is a general result: the combination of relatively low core to envelope contrasts and slow surface rotation rates sets stringent total angular momentum bounds on secondary clump stars, requiring substantial loss. We therefore conclude that if the angular momentum profile of the star is monotonic, then enhanced angular momentum loss must be occurring. It is encouraging that the PMM loss law, not designed for this physical regime, yields reasonable surface rotation rates using a solar-only calibration.}

{Following on this conclusion that enhanced angular momentum loss is occurring, we can then discuss whether any amount of radial differential rotation is required in these stars. Looking at the models that enforce rigid rotation in the whole star (PMMSB), it seems clear that these predict core rotation periods longer than what is observed. This does not rule these models out, however, because there could be differential rotation with depth in the radiative interior. This does suggest, however, that some amount of radial differential rotation could be present in the convective envelope. A maximally differentially rotating convection zone (PMMcz2) predicts core rotation rates significant faster than the observations. We therefore also show in Figure 
\ref{Fig:Mosser} the predicted core rotation rates for a moderately (R$^{-1}$) differentially rotating model with enhanced angular momentum loss (PMMcz1). We suggest that this model matches all of the core and surface rotation constraints better than any of our limiting cases. However, we caution that this is not a fit to the data, that there is no reason the exponent must be an integer, and that these models do not take into account any differential rotation in the radiative interior, which is seen to occur in first ascent giants \citep{DiMauro2016, KlionQuataert2017, Beck2017} and may also be present in core helium burning stars.} 

\section{Conclusions}

Secondary clump stars provide unique insights into stellar rotation, which is both consequential for a wide range of astrophysical problems and challenging to understand theoretically. We find clear evidence that the surface rotation rates of these intermediate mass core helium burning stars are much slower than one would expect from angular momentum conservation and with the adoption of a rigid rotation profile.  This is consistent with prior literature results, on more limited samples.  Our work represents an advance both from the availability of substantially more data and from a confrontation with well-motivated theoretical models. We find that differential rotation in the radiative cores of evolved giants does not significantly impact this conclusion, while the tension between data and theory would be increased if there was significant differential rotation in the main sequence precursors.  We investigated the impact of thermal winds from red giant mass loss, and find that they do not materially impact our predictions. Widely used magnetized wind law prescriptions, based on the \citet{Kawaler1988} formalism, are also found to predict minimal loss levels for these stars. We are therefore left with two (not mutually exclusive) options for reducing the surface rotation rates of these stars: enhanced angular momentum loss in magnetized solar-like winds or the presence of radial differential rotation in the surface convection zones of these stars.

Modern angular momentum loss prescriptions, such as the \citet{vanSadersPinsonneault2013} loss law adopted here, predict significant spin down for these stars, even though the models are calibrated only on the sun and not tuned to reproduce our data directly. We find this to be encouraging evidence that such models are to be preferred for gyrochronology studies. However, models with rigid rotation in the convection zone predict mass trends in the mean rotation rate not visible in the data, {although the data is currently sparse and this conclusion is suggestive rather than definite. This could serve either as evidence for an additional factor, such as differential rotation with depth in the envelope, or a possible defect in the extrapolated mass trends in the torques.} 
 
Differential rotation in the convection zone can also produce slow surface rotation, and some authors \citep[e.g.][]{KissinThompson2015} have argued that asteroseismic data is better fit with uniform rotation in radiative cores and differential rotation in envelopes than the reverse. In our view, the balance of evidence supports the existence of differential rotation in radiative regions of evolved stars \citep{DiMauro2016, KlionQuataert2017} and stars on the upper main sequence \citep{Triana2015}. Our models constructed with differential rotation and weak torques are broadly consistent with the observed surface rotation rates, similar to models with enhanced torques and uniform rotation. However, a full consideration of all constraints makes a solution without enhanced loss unlikely. Low core to envelope contrast ratios \citep[e.g.][]{Deheuvels2015} set strict bounds on the angular momentum content of the convection zone, implying that substantial post-main-sequence loss is essential for reproducing all of the data. This does not rule out more modest differential rotation in convection zones, which would be interesting to quantify as a function of rotation and evolutionary state. Matching both core and surface distributions simultaneously as well as the available trends with mass and surface gravity seems to require both loss and differential rotation. We therefore conclude that both effects are acting in these stars.

We have not considered for this work recent suggestions that loss rates might decrease sharply above some Rossby threshold \citep{vanSaders2016}. This would in general increase the necessary loss or differential rotation rates required to explain the data. We have also made the simplifying assumption for this work that differential rotation turns on instantaneously at the end of the main sequence. If it in fact begins above some Rossby threshold, we would expect a delayed onset of differential rotation for faster rotating stars which might be observable in the Hertzsprung Gap, and could reduce the predicted range of rotation rates in the presence of substantial loss. We have also not included a consideration of binary mass transfer events, which could produce rapid rotators. Such stars are definitely seen in the field \citep{Tayar2015}, but the rate is actually higher in low mass stars than in our targets \citep{Ceillier2017}, as would be expected given the smaller maximum radii of the more massive stars on the red giant branch. 

Measuring stellar surface rotation in evolved stars can be challenging, and particular caution should be employed when interpreting measurements close to detection limits. However, it is encouraging that surface rotation is detectable in the bulk of our targets, suggesting that we are close to the sensitivity needed to fully characterize the sample. More extensive and precise stellar rotation and asteroseismic data could therefore help distinguish between enhanced loss and radial differential rotation in the surface convection zone. Our models with strong torques predict significant spin down during the core helium burning phase, rather than prior to it, which produces a distinct trend in surface gravity distinguishable from expansion and angular momentum conservation. We could also distinguish between the two possibilities if we had a large unbiased sample of stars whose surface rotation rates were reliable down to about 2 \kms\ with well-known masses. It might also be possible to compare measurements of seismic envelope rotation rates to surface rates from spots or rotational broadening to check for strong envelope differential rotation \citep[see] [for a discussion]{DiMauro2016}.

Finally, we note that the processes invoked to change the rotational evolution of stars we discuss here might have significant consequences for angular momentum evolution and gyrochronology. Testing loss rates across a wide range of masses and radii is crucial, and it is important for internal angular momentum transport models to distinguish between core and enveloped differential rotation.

\acknowledgements{We thank Tanda Li and Isabel Colman for confirming that many of the very rapid rotators in the \citet{Ceillier2017} sample were spurious. We thank Tugdual Ceillier, Rafael Garc{\'i}a, Paul Beck, Savita Mathur, and J.J. Hermes for helpful discussions. MP and JT acknowledge support from NASA grant NNX15AF13G.}

\bibliographystyle{apj} 
\bibliography{/home/spitzer/tayar/Documents/Apogee/latex/RapidRottext2.bib}
\end{document}